%% file: ArnalteMur_ALHclustering_v2.tex
\documentclass[useAMS,usenatbib]{mn2e}

\usepackage{graphicx}
\usepackage{amsmath}
\usepackage{amssymb}

\usepackage[pdftex,bookmarks,colorlinks,breaklinks]{hyperref}
\hypersetup{linkcolor=black,citecolor=black,filecolor=black,urlcolor=black} % black links, for printed output

\newcommand{\hMpc}{\, h^{-1} \, \mathrm{Mpc}}
\newcommand{\kmsMpc}{\, \mathrm{km} \, \mathrm{s}^{-1} \, \mathrm{Mpc}^{-1}}
\newcommand{\hMsol}{\, h^{-1} \, \mathrm{M}_{\sun}}

\title[ALHAMBRA: evolution of galaxy clustering]
{The ALHAMBRA survey: evolution of galaxy clustering since $z \sim 1$}

\author[P. Arnalte-Mur et al.]
{
P.~Arnalte-Mur$^{1,\star}$, V.~J.~Mart\'inez$^{2,3,4}$, P.~Norberg$^1$, A.~Fern\'andez-Soto$^{4,5}$, \newauthor
B.~Ascaso$^6$,  A.~I.~Merson$^7$, 
J.~A.~L.~Aguerri$^8$, F.~J.~Castander$^9$, L.~Hurtado-Gil$^{2,5}$, \newauthor C.~L\'opez-Sanjuan$^{10}$,  A.~Molino$^6$, A.~D.~Montero-Dorta$^{11}$, M.~Stefanon$^{12}$, 
E.~Alfaro$^6$, \newauthor T.~Aparicio-Villegas$^{13}$, N.~Ben\'itez$^6$, T.~Broadhurst$^{14}$, J.~Cabrera-Ca\~no$^{15}$, \newauthor J.~Cepa$^{8,16}$, M.~Cervi\~no$^{6,8,16}$,  
D.~Crist\'obal-Hornillos$^{10}$, A.~del~Olmo$^6$, \newauthor R.~M.~Gonz\'alez~Delgado$^6$, C.~Husillos$^6$, 
L.~Infante$^{17}$, I.~M\'arquez$^6$, J.~Masegosa$^6$, \newauthor M.~Moles$^{10}$, J.~Perea$^6$, M.~Povi\'c$^6$, F.~Prada$^6$,  J.~M.~Quintana$^6$
\\
$^1$Institute for Computational Cosmology, Department of Physics, Durham University, South Road, Durham DH1 3LE, UK\\
$^2$Observatori Astron\`omic, Universitat de Val\`encia, C/ Catedr\`atic Jos\'e Beltr\'an 2, E-46980, Paterna, Spain\\
$^3$Departament d'Astronomia i Astrof\'isica, Universitat de Val\`encia, E-46100, Burjassot, Spain\\
$^4$Unidad Asociada Observatorio Astron\'omico (IFCA-UV), E-46980, Paterna, Spain\\
$^5$Instituto de F\'isica de Cantabria (CSIC-UC), E-39005 Santander, Spain\\
$^6$IAA-CSIC, Glorieta de la Astronom\'ia s/n, 18008 Granada, Spain\\
$^7$Department of Physics and Astronomy, University College London, Gower Street, London WC1E 6BT, UK\\
$^8$Instituto de Astrof\'isica de Canarias, V\'ia L\'actea s/n, 38200 La Laguna, Tenerife, Spain\\
$^9$Institut de Ci\`encies de l'Espai (IEEC-CSIC), Facultat de Ci\`encies, Campus UAB, 08193 Bellaterra, Spain\\
$^{10}$Centro de Estudios de F\'isica del Cosmos de Arag\'on, Plaza San Juan 1, 44001 Teruel, Spain\\
$^{11}$Department of Physics and Astronomy, University of Utah, Salt Lake City, UT 84112, USA\\
$^{12}$Physics and Astronomy Department, University of Missouri, Columbia, MO 65211, USA\\
$^{13}$Observat\'orio Nacional-MCT, Rua Jos\'e Cristino, 77. CEP 20921-400, Rio de Janeiro-RJ, Brazil\\
$^{14}$Department of Theoretical Physics, University of the Basque Country UPV/EHU, 48080 Bilbao, Spain\\
$^{15}$Departamento de F\'isica At\'omica, Molecular y Nuclear, Facultad de F\'isica, Universidad de Sevilla, 41012 Sevilla, Spain\\
$^{16}$Departamento de Astrof\'isica, Facultad de F\'isica, Universidad de La Laguna, 38206 La Laguna, Spain\\
$^{17}$Departamento de Astronom\'ia, Pontificia Universidad Cat\'olica. 782-0436 Santiago, Chile\\
$^{\star}$ E-mail: pablo.arnalte-mur@durham.ac.uk
}

\begin{document}

\date{Accepted by MNRAS, 2014 April 4.}

\pagerange{\pageref{firstpage}--\pageref{lastpage}} \pubyear{2014}

\maketitle

\label{firstpage}

\begin{abstract}
We study the clustering of galaxies as function of luminosity and redshift in the range $0.35 < z < 1.25$ using data from the Advanced Large Homogeneous Area Medium Band Redshift Astronomical (ALHAMBRA) survey.
The ALHAMBRA data used in this work cover $2.38 \deg^2$ in 7 independent fields, after applying a detailed angular selection mask, with accurate photometric redshifts, $\sigma_z \lesssim 0.014 (1+z)$, down to $I_{\rm AB} < 24$.
Given the depth of the survey, we select samples in $B$-band luminosity down to $L^{\rm th} \simeq 0.16 L^{*}$ at $z = 0.9$.
We measure the real-space clustering using the projected correlation function, accounting for photometric redshifts uncertainties. 
We infer the galaxy bias, and study its evolution with luminosity.
We study the effect of sample variance, and confirm earlier results that the COSMOS and ELAIS-N1 fields are dominated by the presence of large structures.
For the intermediate and bright samples, $L^{\rm med} \gtrsim 0.6L^{*}$, we obtain a strong dependence of bias on luminosity, in agreement with previous results at similar redshift.
We are able to extend this study to fainter luminosities, where we obtain an almost flat relation, similar to that observed at low redshift.
Regarding the evolution of bias with redshift, our results suggest that the different galaxy populations studied reside in haloes covering a range in mass between $\log_{10}[M_{\rm h}/(\hMsol)] \gtrsim 11.5$ for samples with $L^{\rm med} \simeq 0.3 L^{*}$ and $\log_{10}[M_{\rm h}/(\hMsol)] \gtrsim 13.0$ for samples with $L^{\rm med} \simeq 2 L^{*}$, with typical occupation numbers in the range of $\sim 1 - 3$ galaxies per halo.
\end{abstract}

%%%%%%%%%%%%%%%%%%%%%%%%%%%%%%%%%%%%%%%%%%%%%%

\section{Introduction}
\label{sec:intro}

The large-scale structure (LSS) of the Universe is one of the main observables that we can use to obtain information about the nature of dark matter and cosmic acceleration.
The simplest way to study the LSS is to study the spatial distribution of galaxies in surveys covering cosmologically significant volumes.
Although the galaxy distribution is closely related to the global matter distribution, they are not equal.
The relation between both distributions is known as galaxy bias, and it depends on the processes of galaxy formation and evolution.
In the simplest case, one can consider the galaxy contrast to be proportional to the matter contrast. 
Then, the bias is simply the constant of proportionality, which is independent of scale.
Being able to understand and model this bias is crucial for the correct interpretation of the cosmological information that can be obtained from the analysis of galaxy clustering.

As the bias encodes information about the galaxy formation and evolution process, it is logical to expect that it will be different for different galaxy populations.
In other words, the clustering properties of galaxies should depend on some of their intrinsic properties, such as stellar mass, star formation rate or age, and should evolve with time.
This phenomenon, known as galaxy segregation, is observed when studying the dependence of clustering on different observables such as luminosity, colour, or morphology.
In general, it is observed that bright, red, elliptical galaxies are more strongly clustered (i.e., they have a larger bias) than faint, blue, spiral ones \citep[see e.g.][]{dav76a,ham88a,mad03a,ski08a,mar10a,zeh10a}. 

In this work, we focus on the dependence of the galaxy bias on luminosity, and the evolution of this relation with redshift.
This dependence has been studied extensively in the local Universe using both the Two-degree Field Galaxy Redshift Survey \citep[2dFGRS,][]{nor01a,nor02b} and Sloan Digital Sky Survey \citep[SDSS,][]{teg04b,zeh05a,zeh10a}.
\citet{Guo2013a} also studied this relationship at $z \sim 0.5$ using data from the Baryon Oscillation Spectroscopic Survey (BOSS).
The bias shows a weak dependence on luminosity $L$ for galaxies with $L < L^{*}$, where $L^{*}$ is the characteristic luminosity parameter of the Schechter function.
For $L \gtrsim L^{*}$, however, this relation steepens, and the bias clearly increases with luminosity.

These studies of galaxy clustering and luminosity segregation have been extended to redshifts in the range $z \sim 0.5 - 1$, using state-of-the-art spectroscopic surveys, such as the VIMOS-VLT Deep Survey \citep[VVDS,][]{pol06a,abb10a}, the Deep Extragalactic Evolutionary Probe survey \citep[DEEP2,][]{coi06a, coi08a}, the zCOSMOS survey \citep{men09a} or the VIMOS Public Extragalactic Redshift Survey \citep[VIPERS,][]{mar13a}, or photometric surveys such as the Canada-France-Hawaii Legacy Survey (CFHTLS) Wide survey \citep{McCracken2008a,Coupon2012a}.
Recently, \citet{Skibba2013c} used an intermediate method, somehow similar to the one presented in this work.
They used low-resolution spectroscopy data (with a typical redshift precision of $\sigma_z/(1+z) = 0.005$) from the PRIsm Multi-object Survey (PRIMUS) to study galaxy clustering in the range $0.2 < z < 1$.
Overall, these studies show strong evidence for luminosity segregation at these redshifts, with the relation between bias and luminosity being slightly steeper in this case than in local studies.
However, the luminosity range covered by these surveys is more limited in these cases, and is restricted typically to $L^{\rm th} \gtrsim 0.3 L^{*}$.

The presence of this bias parameter can be understood in a natural way in the context of the halo model \citep[e.g.][]{sel00a,pea00a,coo02a}.
In this model, the matter distribution is decomposed into a population of massive virialised dark matter haloes that form at the peaks of the density field, and galaxies form within these haloes.
The bias parameter for dark matter haloes can be modelled, and depends on the properties of the halo such as its mass \citep[e.g.][]{Sheth2001i,mo02a}.
Studying the clustering of a certain population of galaxies gives therefore information on the characteristics of the haloes that host them.
In this context, luminosity segregation indicates that more luminous galaxies form preferentially in more massive haloes than fainter ones.

In this work, we use data from the  Advanced Large Homogeneous Area Medium-Band Redshift Astronomical (ALHAMBRA) survey \citep{mol08a,Molino2013a} to study galaxy clustering and luminosity segregation for redshifts in the range $0.35 < z < 1.25$, using the two-point correlation function.
ALHAMBRA is a deep photometric survey, which uses a total of 23 optical and near-infrared (NIR) bands in order to obtain accurate and reliable photometric redshifts (photo-$z$) for a large number of objects, in a nominal area of $4~\deg^2$ over 8 independent fields.
It is therefore well suited to study the large-scale distribution of galaxies in significant volumes over this redshift range, providing an opportunity to explore the clustering of fainter galaxies than it is possible using spectroscopic surveys.
Moreover, the use of several independent fields allows us to use ALHAMBRA to study the effect of sample variance in the clustering measurements, and in particular the effect of large structures present in the samples.
\citet{lopez2013a} also exploited this independence of the ALHAMBRA fields to study the effect of sample variance on merger fraction studies.

In Section~\ref{sec:data} we present the ALHAMBRA data used in this work (characterised in more detail in \citealt{Molino2013a}), and our selection of samples. 
We also present here the mock catalogues created to test our clustering methods.
In Section~\ref{sec:selfunc} we explain how we model the selection function of the survey, and in particular the masks created to reproduce the angular selection.
Section~\ref{sec:projcf} presents our method to estimate the projected correlation function (a real space quantity) taking into account the effect of the photometric redshifts, following \citet{arn09a}. We also present our error estimation method, and leave for Appendix~\ref{sec:pimax} the detailed justification of our line-of-sight integration limit.
Our results are presented in Section~\ref{sec:results}.
We show the correlation functions obtained for our different samples, including the modelling in terms of a simple power law model (Sect.~\ref{ssec:powlaw}), and of a $\Lambda$ cold dark matter ($\Lambda$CDM) model in order to derive the bias parameter (Sect.~\ref{ssec:bias}).
We also make use of the independence of the surveyed fields to study the effect of sample variance on our measurements (Sect.~\ref{sec:cvariance}), and compare our results with those of previous surveys in a similar redshift range (Sect.~\ref{sec:othersurveys}).
In Appendix~\ref{sec:systematics} we present the tests done using the mock catalogues to test the reliability of the results, and Appendix~\ref{sec:numer-results} contains numerical tables of our results.
Finally, in Sect.~\ref{sec:conc} we discuss our results and summarise our conclusions.

Unless noted otherwise, we use a fiducial flat $\Lambda$CDM cosmological model with parameters $\Omega_{\rm M} = 0.27$, $\Omega_{\Lambda} = 0.73$, $\Omega_{\rm b} = 0.0458$, and $\sigma_8 = 0.816$ based on the WMAP7 results \citep{kom11a}. 
All the distances used are co-moving, and are expressed in terms of the Hubble parameter $h \equiv H_0/100 \kmsMpc$.
Absolute magnitudes are given as $M - 5 \log_{10}(h)$, even when not explicitly indicated.

%%%%%%%%%%%%%%%%%%%%%%%%%%%%%%%%%%%%%%%%%%%%%%
\section{Data used: the ALHAMBRA Survey}
\label{sec:data}

The ALHAMBRA survey \citep{mol08a} is a photometric survey which covers a total of $4 \deg^2$ in the sky, using 20 medium-band filters in the optical range, and three standard broad-band filters ($J$, $H$, and $K_s$) in the NIR.
The survey was carried out using the 3.5-m telescope at the Centro Astron\'omico Hispano-Alem\'an (CAHA)\footnote{http://www.caha.es} in Calar Alto (Almer\'{\i}a, Spain). 
The camera used for the optical observations was the Large Area Imager for Calar Alto (LAICA)\footnote{http://www.caha.es/CAHA/Instruments/LAICA}, and Omega-2000\footnote{http://www.caha.es/CAHA/Instruments/O2000} was used for the NIR observations.

The optical filter system for the ALHAMBRA survey was specifically designed to optimise the output of the survey in terms of photo-$z$ accuracy and number of objects with reliable $z$ determination \citep{ben09a}. 
It consists of a set of 20 contiguous, equal-width, medium-band filters of width $FWHM~\simeq~310$~\AA\ covering the full optical range, between $3500$ and $9700$ \AA\ \citep{apa10a}. 
The survey is complemented by observations in the standard NIR filters $J$, $H$ and $K_s$. 
The homogeneous spectral coverage of this system minimises the variations in the selection functions of the different objects with redshift.
The NIR observations help eliminate some degeneracies in the photo-$z$ determination while at the same time improving the determination of important galaxy properties such as stellar mass.

\subsection{ALHAMBRA imaging data}
\label{ssec:imaging-data}

\begin{table}
  \centering
\caption{Properties of the seven ALHAMBRA fields used in this work. 
We list the number of frames $N_{\rm f}$ included in the current catalogue in each case (where a completed field corresponds to 8 frames), the area $A_{\rm eff}$ covered by the survey according to our angular selection mask, the number of galaxies $N_{\rm g}$ included in the catalogue used (at $I < 24$), and the resulting surface number density $N_{\rm g}/A_{\rm eff}$.
  We also list other surveys which have overlap with each of the fields, see \citet{mol08a} for details.}
  \label{tab:fields}
  \begin{tabular}{lcccc}
    \hline
    Field      &    $N_{\rm f}$     & $A_{\rm eff}$  & $N_{\rm g}$    & $N_{\rm g}/A_{\rm eff}$\\
               &                  & $(\deg^2)$   &               & $(\deg^{-2})$ \\
    \hline
    ALH-2/DEEP2     &    8         & 0.377       & 26759         & 70979 \\ 
    ALH-3/SDSS      &    8         & 0.404       & 28331         & 70126\\
    ALH-4/COSMOS    &    4         & 0.203       & 16877         & 83138\\
    ALH-5/HDF-N     &    4         & 0.216       & 16629         & 76986\\
    ALH-6/GROTH     &    8         & 0.400       & 28892         & 72230\\
    ALH-7/ELAIS-N1  &    8         & 0.406       & 29530         & 72734\\
    ALH-8/SDSS      &    8         & 0.375       & 27615         & 73640\\
    \hline
    Total           &   48        &  2.381       & 174633        & 73344\\
    \hline
  \end{tabular}
\end{table}

The data used in this work correspond to the photometric catalogue described in \citet{Molino2013a}\footnote{This ALHAMBRA catalogue is publicly available at http://www.alhambrasurvey.com/}.
It contains data for a nominal area of $3 \deg^2$ distributed over 7 fields in the sky, in order to minimise the effects of sample variance (see Table~\ref{tab:fields}).
The minimum distance between fields is $17^{\circ}$, so we can safely consider them as independent.
The fields were primarily chosen because of their low extinction, and trying to have significant overlap with other surveys \citep{mol08a}.
Each field is typically composed of 8 frames forming two strips of $\sim 15' \times 1^{\circ}$, separated by a $\sim 15'$ gap.
We discuss in detail the geometry of the different fields in Section~\ref{ssec:angmask}.
We developed our own pipelines for the reduction of the imaging data, including bias, flatfield and fringing corrections.
The details of the data reduction can be found in \citet[][in prep.]{cri09a} for both the optical and NIR data.

The detection of objects for inclusion in the catalogue is performed in synthetic images built using a combination of the ALHAMBRA filters in the range $7000 < \lambda < 9700$ \AA\ to match the Hubble Space Telescope $F814W$ filter (hereafter denoted as our $I$ band).
Matched photometry is then obtained for these detected objects in the 23 ALHAMBRA filters.
We restrict our analysis to the magnitude range $I < 24$, where the catalogue is photometrically complete and we do not expect any significant field-to-field variation in the depth \citep[see section 3.8 in][]{Molino2013a}.

We eliminate stars from the catalogue using the star-galaxy separation method described in \citet{Molino2013a}, which uses information on both the geometry and colours of the sources. 
In particular, we use the stellar flag given in the catalogue, and select only objects with \textsc{Stellar Flag}~$< 0.7$.
This method is only reliable for $I < 22.5$.
However, at $I=22.5$ the fraction of stars in the sample is $\sim 1 \%$, and we expect it to decrease at fainter magnitudes.
Therefore the possible effect of stellar contamination at $I > 22.5$ is negligible.
The final catalogue used contains a total of $N_{\rm T} = 174,633$ galaxies.

\subsection{Photometric redshifts}
\label{ssec:photo-z}

\begin{figure}
  \centering
  \includegraphics[width=\columnwidth]{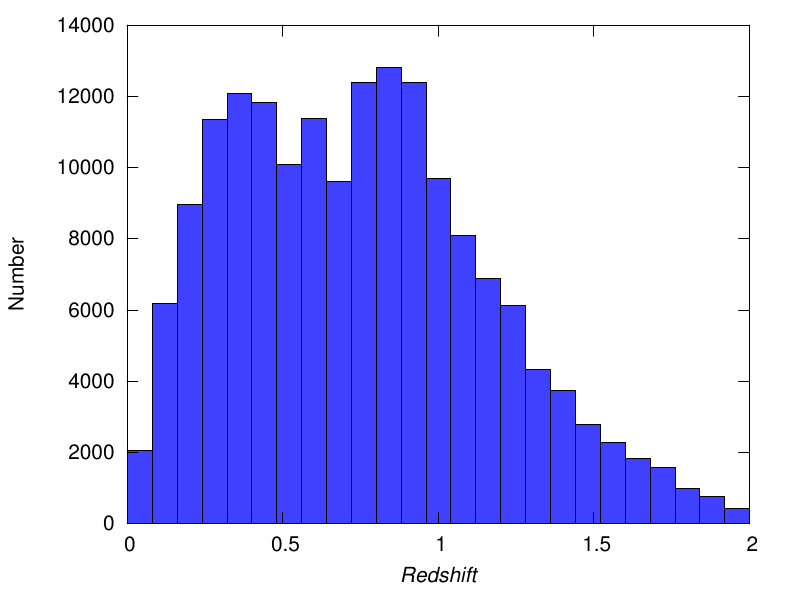}
  \caption{Redshift distribution of the $174,633$ galaxies in the ALHAMBRA catalogue used in this work. The distribution shown corresponds directly to a histogram of the `best' photo-$z$ for each galaxy, in bins of width 0.08.}
  \label{fig:zhist}
\end{figure}

Photometric redshifts were estimated for this catalogue using an updated version of the Bayesian Photometric Redshift (\textsc{BPZ}) code \citep{ben00a}, including a new prior and spectral template library (Ben\'itez et al., in prep.), and a new technique for the re-calibration of the photometric zero points.
\citet{Molino2013a} discussed in detail the methods used for the redshift estimation and the characteristics of the photo-$z$ obtained. 
They performed a comparison for the $\sim 7000$ galaxies with measured spectroscopic redshift (see their figure 25) and showed that the global accuracy in the photo-$z$ is $\sigma_z \lesssim 0.014 (1+z)$ for $I < 24.5$.
We show the distribution of photo-$z$ for this catalogue in Fig.~\ref{fig:zhist}. 
The median redshift of the catalogue is $z_{\rm med} = 0.75$,  with the bulk of the redshift distribution in the range $0.35 < z < 1.25$ that we study in this work.

\begin{figure}
  \centering
  \includegraphics[width=\columnwidth]{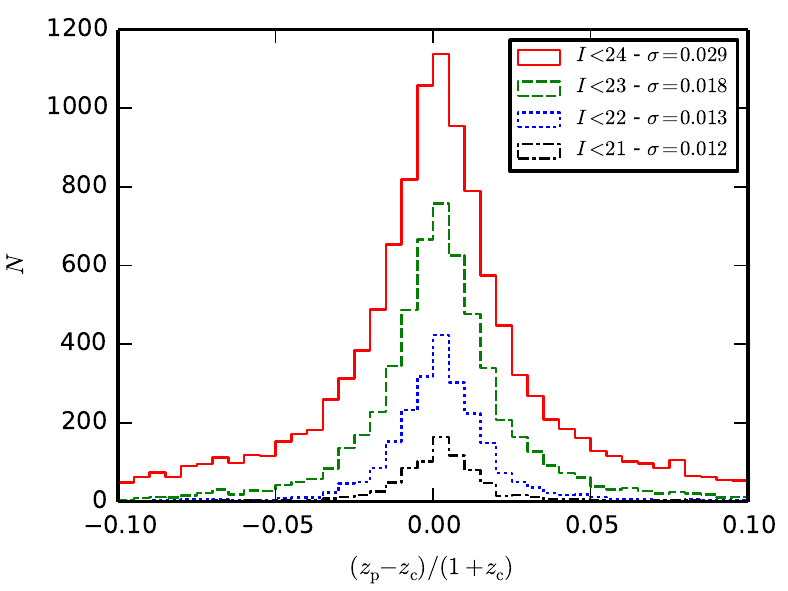}
  \caption{Distribution of the relative differences between the ALHAMBRA photometric redshift ($z_{\rm p}$) and the COSMOS photometric redshift ($z_{\rm c}$) for the objects matched between the two catalogues. We show this distribution for different $I$-band magnitude selections, as shown in the label. We quote in each case the dispersion $\sigma$ estimated using the NMAD method.}
  \label{fig:hist_cosmos}
\end{figure}

As an additional test of the reliability of the photometric redshifts used, we made a comparison with the Cosmic Evolution Survey (COSMOS) photo-$z$ catalogue described in \citet{ilb08a}.
This catalogue contains photometric redshift determinations with comparable accuracy and depth to those in the ALHAMBRA catalogue, and overlaps with the field ALH-4 (see Table~\ref{tab:fields}).
We matched both catalogues using a separation radius of $1''$ in the angular position, and obtained a sample of 12832 objects common to both catalogues. 
We show the distribution of the relative redshift differences for this sample in Fig.~\ref{fig:hist_cosmos}, where we also quote the dispersion in the results measured using the normalised median absolute deviation (NMAD) method \citep[see e.g.][]{bra08a}.

We compare the dispersion obtained in this way to a simple estimate based on the redshift errors quoted in both catalogues. 
In each case, we estimate the typical redshift uncertainty (in both ALHAMBRA and COSMOS) as the mean of the $1\sigma$ errors quoted for each object in the sample.
Our estimate for the dispersion in the difference shown in Fig.~\ref{fig:hist_cosmos} is then $\sigma_{\rm diff} = \sqrt{\sigma_{\rm ALH}^2 + \sigma_{\rm COS}^2}$.
We obtain that this value of the dispersion obtained from the quoted errors is a good estimate of that observed.
However, for our faintest samples ($I > 23$), we need to increase this estimate by a factor of $\sim1.3$, suggesting that the photo-$z$ uncertainty could be slightly underestimated for these galaxies in both ALHAMBRA and COSMOS.
Hereafter, we quote the error estimates for our samples (e.g. in Table~\ref{tab:samples}) as the mean of the quoted \textsc{BPZ} error for the objects in the sample.
For consistency, we correct this value by the factor of $1.3$ for all samples, although we only see an indication for the underestimation of the errors at the faintest ones.

\subsection{Selection of samples in redshift and luminosity}
\label{ssec:samples}

To study the dependence of clustering properties on both luminosity and cosmic time, we build a series of subsamples splitting the catalogue in redshift and absolute magnitude.

The size of the redshift bins has to be larger than the distance we will integrate over the radial direction, $\pi_{\rm max}$. 
As shown in \citet{arn09a}, using smaller bins may introduce systematic effects in the correlation functions we want to measure. 
Taking this fact into account, and the limitations in volume covered and galaxy density, we decided to use the four redshift bins 
$0.35 < z_{\rm p} < 0.65$, 
$0.55 < z_{\rm p} < 0.85$, 
$0.75 < z_{\rm p} < 1.05$, 
$0.95 < z_{\rm p} < 1.25$.
We allow for overlap between consecutive bins in order to better trace the redshift evolution in our analysis, but one should bear in mind that results for different bins will be therefore correlated.
Our low redshift limit $z_{\rm p} = 0.35$ was set in order for the scales of interest to be well sampled given the angular size of the fields.
At this redshift, the typical size of a field, $1^{\circ}$, corresponds to a projected comoving separation of $17 \hMpc$.
We fixed our high redshift limit at $z_{\rm p} = 1.25$ as, for higher redshifts, the quality of the photo-$z$ and the number density of objects are significantly reduced.

In addition to the redshift selection, we also apply a set of cuts in the rest-frame $B$-band absolute magnitude $M_B$. 
We use this band for the selection as the region of the spectrum corresponding to it is well sampled by the ALHAMBRA filters (including the NIR filters) for the whole redshift range studied.
Moreover, as this same band (or similar ones as $g$) is used for luminosity selection by other surveys at these redshifts, this will allow for more direct comparisons.
The $M_B$ for each object is obtained as a by-product of the photo-$z$ estimation, and includes the appropriate $K$-correction at the best value of $z_{\rm p}$.
We use `threshold samples', meaning that we will impose a faint luminosity threshold, but not a bright limit. 
In this way, we obtain approximately volume-limited samples, but also we can study the luminosity dependence of clustering, and its evolution.
Following \citet{men09a} and \citet{abb10a}, we apply an absolute magnitude threshold depending linearly on redshift as 
\begin{equation}
\label{eq:p1alh:Mthevol}
M_B^{\rm th}(z) = M_B^{\rm th}(0) + A z_{\rm p} \,,
\end{equation} 
in order to follow the evolution of samples corresponding approximately to the same galaxy population. 
The value of the constant $A$ characterises the typical luminosity evolution of the galaxies in the catalogue.
We use here a value of $A = -0.6$, which we selected to produce samples with similar number density across the whole redshift range.
This value is also similar to the observed evolution of the typical luminosity parameter $M^{*}$ derived from luminosity function studies at similar redshifts \citep{ilb05a,zuc09a}.

\begin{figure}
  \centering
  \includegraphics[width=\columnwidth]{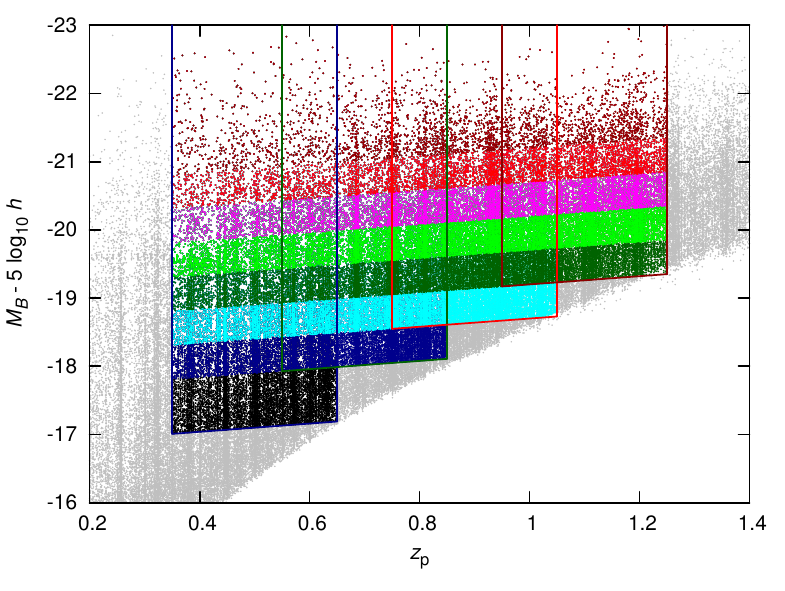}
  \caption{Selection of samples in the absolute $B$-band magnitude $M_B$ vs. photometric redshift diagram. 
    The different coloured dots show the eight magnitude cuts, while the lines mark the boundaries of our redshift bins. 
    See the main text and Table~\ref{tab:samples} for the details of the sample selection.}
  \label{fig:zMB}
\end{figure}

\begin{table*}
  \centering
  \caption{Characteristics of the different samples selected in redshift and luminosity. For each sample we quote the redshift range, $B$-band absolute magnitude threshold at $z=0$ $M_B^{\rm th}(0)$ (see equation~\ref{eq:p1alh:Mthevol}), number of galaxies $N$, mean number density $\bar{n}$, median redshift $z_{\rm med}$, median absolute magnitude $M_B^{\rm med}$, median luminosity $L^{\rm med}$ as function of $L^{*}(z_{\rm med})$, typical redshift error $\sigma_z/(1+z)$, and typical line-of-sight distance error $r(\sigma_z)$ (see the text for details).}
  \label{tab:samples}
  \hspace*{-1cm}\input{ArnalteMur_table_2.tex}
\end{table*}

We show in Fig.~\ref{fig:zMB} the actual cuts made in the redshift -- absolute magnitude plane to define our samples, and list the properties of all the samples used in Table~\ref{tab:samples}.
We estimate the error in the mean number density $\bar{n}$ of each sample using a block bootstrap method based on the 7 independent fields.
For each sample, we compute the typical $z_{\rm p}$ error $\sigma_z/(1+z)$ as described in Sect.~\ref{ssec:photo-z}, and the line-of-sight distance that corresponds to this uncertainty, $r(\sigma_z)$, measured at the median redshift $z_{\rm med}$ of the sample.
We also measure the median absolute luminosity, $M_B^{\rm med}$, and express it in terms of the typical luminosity parameter $L^{*}$ at $z_{\rm med}$.
We compute $L^{*}(z)$ from a linear fit to the results of \citet{ilb05a}.

\subsection{Mock catalogues}
\label{ssec:mock-catalogues}

To test our methods for clustering and error estimation, and to provide a test bench for future ALHAMBRA studies, we use a set of mock catalogues, based on the Millennium dark matter simulation \citep{spr05a}.
We populate the dark matter haloes with galaxies using the \citet{Lagos2011a} version of the semi-analytic galaxy formation model \textsc{Galform} \citep{Cole2000a}.
In addition to other physical parameters, we compute the photometry for each of the galaxies in the model using the 24 ALHAMBRA filters, including the synthetic $I$ band and, for completeness, also using the five SDSS broad-band filters $ugriz$.
A light-cone is built from the simulation's snapshots up to $z = 2$, reproducing the photometric depth of the survey.
In order to properly model the evolution of structures along the line of sight, the galaxy positions are interpolated between snapshots.
The procedure used to generate the light-cone mocks is presented in detail in \citet{mer13a}.
The cosmological model used for the mocks is set by that of the Millennium simulation, which uses the parameters $\Omega_{\rm M} = 0.25$, $\Omega_{\Lambda} = 0.75$, $\sigma_8 = 0.9$.
We will use these parameters when doing tests with the mocks in Appendices~\ref{sec:analyt-model-determ} and \ref{sec:systematics}.

We generate a $200 \deg^2$ light-cone, which is divided in 50 non-overlapping mock ALHAMBRA realisations.
Each of these realisations reproduces the ideal geometry of the full survey, containing 8 fields covering $0.5 \deg^2$ each, for a total of $4\deg^2$ per realisation.
The fields in each realisation are as separated as possible within our light-cone geometry.
Each field is formed by two strips of $15' \times 1^{\circ}$, separated by a $15'$ gap, approximately reproducing the geometry of the ALHAMBRA fields, as described in Sect.~\ref{ssec:angmask}.

To simulate the photometric redshifts for the galaxies in the mock we proceeded as follows. 
We first use the original rest-frame photometry and spectroscopic redshifts in the mock to assign to each galaxy a spectral type from the same \textsc{BPZ} template library used to estimate photo-$z$ in the real data\footnote{We do this assignment running \textsc{BPZ} with the \textsc{Only\_Type} option}.
Then, we measure consistent ALHAMBRA photometry for these spectral types by using the ALHAMBRA filter curve response. 
Finally, we estimate the photometric redshifts, together with the spectral types and absolute magnitudes associated with the previous photometry, by running \textsc{BPZ} in normal mode. 
These photometric redshifts are found to be very realistic as their performance is very similar to those obtained for real data, although with a somewhat larger uncertainty ($\sim 30\%$).
All the details can be found in Ascaso et. al (in prep.).

%%%%%%%%%%%%%%%%%%%%%%%%%%%%%%%%%%%%%%%%%%%%%%
\section{Modelling the selection function}
\label{sec:selfunc}

To study the clustering of the galaxies in a survey, it is crucial to understand and to model its selection function.
In this work,  we separate the angular and radial parts of the selection function, with our angular selection function (or `mask')  defining the geometry of the survey on the sky.
We assume a uniform depth inside the mask, as the catalogue considered does not reach the photometric limit of the survey.

\subsection{Angular selection mask}
\label{ssec:angmask}

\begin{figure}
  \centering
  \includegraphics[width=\columnwidth]{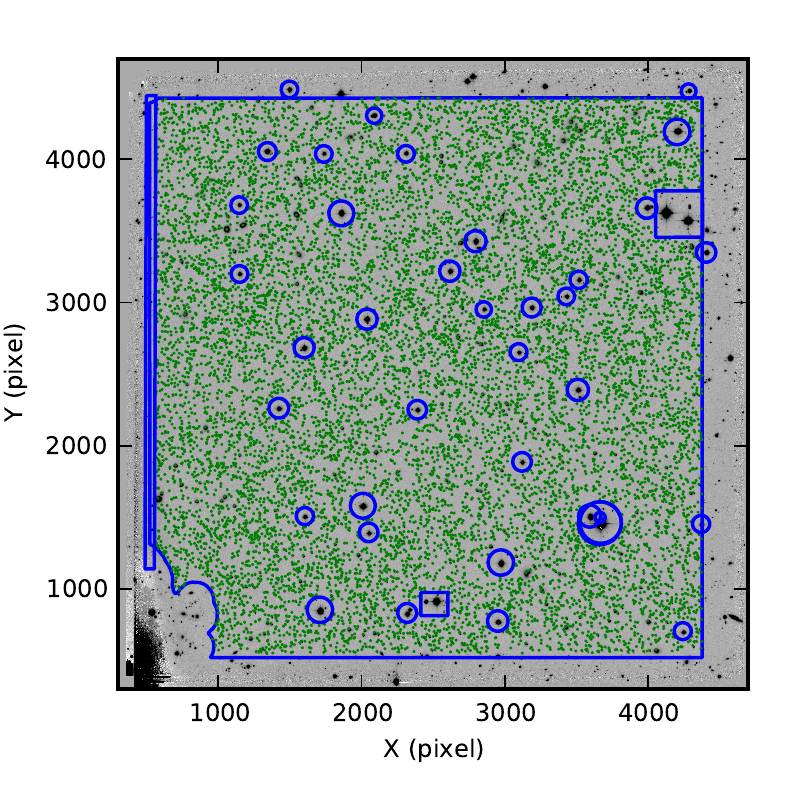} \hfill
  \includegraphics[width=\columnwidth]{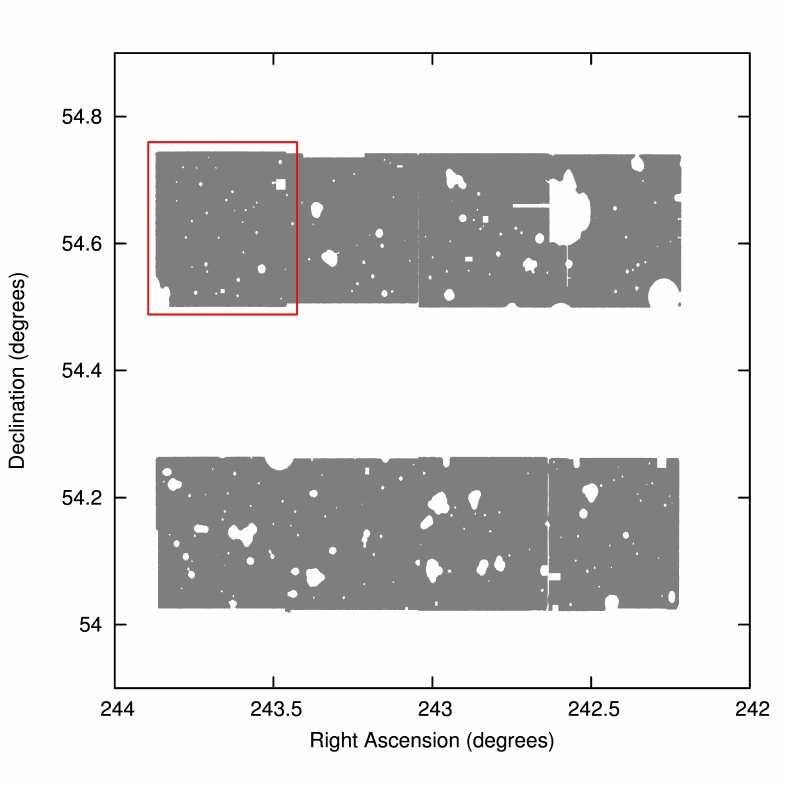}
  \caption{Illustration of the ALHAMBRA angular mask for field ALH-7. Top: synthetic $I$-band image for one of the 8 frames in the field, showing an area of $\sim 16' \times 16'$. Green dots mark the position of the objects included in the catalogue, and the blue lines show the limits of the angular selection mask. Bottom: angular mask for the ALH-7 field. The shaded area corresponds to the regions of the survey that are included in the calculations. The red rectangle marks the area shown in the top image.}
  \label{fig:masks}
\end{figure}

The angular selection mask is defined in the first instance by the coverage of the ALHAMBRA survey.
It consists of independent fields of $\sim 0.5 \deg^2$ each, with a specific geometry set by the configuration of the detectors in the optical camera used, LAICA. 
The camera has four $15.5' \times 15.5'$ detectors, distributed in a square leaving a space of $13.6'$ between them. 
Each of the ALHAMBRA fields consists of two pointings made with this configuration, resulting in two strips of $15.5' \times 58.5'$ with a gap of $13.6'$ between them (see bottom panel of Fig.~\ref{fig:masks}).
For this work, fields ALH-4 and ALH-5 correspond to only one pointing each, and thus are formed by four disjoint $15.5' \times 15.5'$ frames.

Based on that geometry, we define a set of masks describing the sky area which has been reliably observed. 
We start with the \emph{flag images} described in \citet{Molino2013a}, that give information on the areas in which the detection of the objects in the synthetic $I$-band images was performed.
They exclude areas with low exposure time (less than $60\%$ of the maximum in each frame), which mainly correspond to regions next to the borders of each frame, or corresponding to large saturated stars.

To avoid possible variations in depth, which could potentially introduce a spurious clustering pattern, we remove some additional regions from the survey area, taking a conservative approach.
We mask out regions around bright stars, using the Tycho-2 catalogue \citep{Hog2000d}.
The masked regions are circles of radius $33\, \mathrm{arcsec}$ centred on each star.
For the brightest stars ($V < 11$), we extend this radius to $111\, \mathrm{arcsec}$.
We define these radii by observing the typical maximum extension of the stellar haloes in the $I$-band detection images.
Furthermore, we select objects showing saturated detections in the ALHAMBRA catalogues (using the \textsc{Satur\_Flag} parameter, see \citealt{Molino2013a} for details), and mask a region around each of them with a radius twice that of the object itself.

Finally, we mask by hand some obvious defects in the image (typically extended stellar spikes), and some small overlap between contiguous frames.
The latter is needed to avoid double-counting objects from the overlap regions when computing the clustering for the combined field.
To avoid position-dependent differences in the photo-$z$ quality we mask by hand regions which present bad photometric quality in at least 3 of the ALHAMBRA bands (but not necessarily in the $I$ band used for detection). 
This uses the \textsc{irms\_opt\_flag} and \textsc{irms\_nir\_flag} parameters in the catalogue \citep[see][for details]{Molino2013a}.

We defined and combined the different masks using the \textsc{Mangle}\footnote{http://space.mit.edu/$\sim$molly/mangle/} software \citep{ham04a,swa08a}, which allows for an easy manipulation of angular masks, and for some additional routines like generating random catalogues.
These angular masks will be publicly available from http://www.alhambrasurvey.com/.
Fig.~\ref{fig:masks} illustrates the resulting mask for ALH-7.

The total effective area  after applying this mask is $A_{\rm eff} = 2.381 \deg^2$, distributed over the different fields as shown in Table~\ref{tab:fields}. 
Overall, this procedure masks an additional $\sim 15\%$ of the area not yet masked by the original \emph{flag images}. 
This explains the difference in area between this work and \citet{Molino2013a}.

\subsection{Radial selection function}
\label{ssec:radsfunc}

\begin{figure}
  \centering
  \includegraphics[width=\columnwidth]{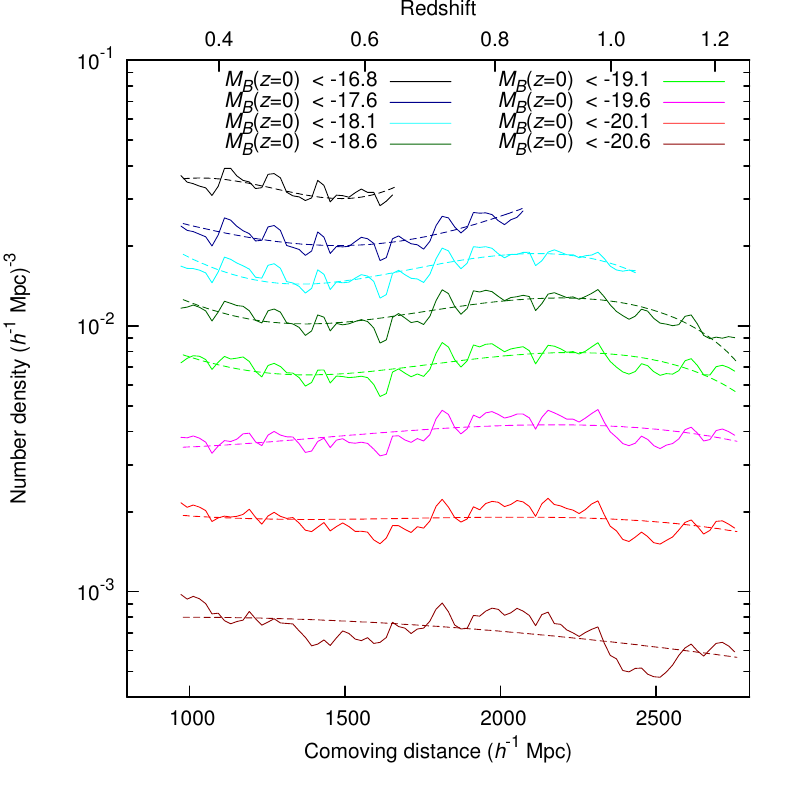}
  \caption{Number density as function of comoving distance (or, equivalently, redshift) for our different cuts in absolute magnitude. We show the function directly measured from the data with a smoothing length of $200 \hMpc$ (continuous lines) and our third-order polynomial fit (dashed lines) in each case.
Lines from top to bottom correspond to samples with fainter to brighter luminosity cuts.}
  \label{fig:radialdens}
\end{figure}

We model the radial selection function for our different samples directly using the observed number density of galaxies as function of comoving distance (or, equivalently, redshift), $n(d)$.
We show in Fig.~\ref{fig:radialdens} the number density of our different samples selected in luminosity (solid lines), measured using a smoothing length of $200 \hMpc$.
Given our redshift-dependent luminosity cut, the number density for each of the samples is approximately constant over the redshift range considered, as expected for nearly volume-limited samples.

However, apart from the small-scale variations due to the presence of structures, we observe some long-range variations in $n(d)$.
We assume the latter are part of our selection function, and model them by fitting a third-order polynomial to $n(d)$ over the full range spanned by each of the samples.
This model is smooth enough not to include possible variations in $n(d)$ due to large-scale structures, to prevent a systematic underestimation of the clustering signal.

We use this smooth model for our clustering measurements as described below. 
However, we performed some tests assuming either a model with constant $n(d)$, or using directly the measured $n(d)$ as our radial selection.
Our results do not change significantly in either case.

One particularity of the radial density of ALHAMBRA as measured here is the presence of a series of regularly spaced `peaks'. 
They can be seen more clearly in Fig.~\ref{fig:radialdens} for the fainter samples (higher $n$), or as a series of vertical `strips' in the distribution of galaxies in Fig.~\ref{fig:zMB}.
The presence of these peaks is the consequence of using only the best value $z_{\rm p}$ of the photometric redshift estimate for each galaxy, instead of the full probability density function $p(z)$ \citep{ben00a}.
We tested whether this issue could introduce any systematic bias in our measurements by creating a new `realisation' of the photometric redshifts: we assigned to each galaxy a new value of $z_{\rm p}$ drawn from a Gaussian distribution centred at the original value, and with a width given by the quoted error. 
Additionally, we randomly selected $5\%$ of galaxies to be `outliers', and assigned them a random value of $z_{\rm p}$ within the studied range.
We computed the projected correlation function for our samples using this new `realisation', and obtained only small changes contained within the quoted errors.
We therefore conclude that the presence of these peaks in $n(d)$ does not significantly bias our results.

%%%%%%%%%%%%%%%%%%%%%%%%%%%%%%%%%%%%%%%%%%%%%%
\section{The projected correlation function calculation in photometric redshift catalogues}
\label{sec:projcf}

The two-point correlation function $\xi(\mathbf{r})$ measures the excess probability of finding two points separated by a vector~$\mathbf{r}$ compared to that probability in a homogeneous Poisson sample \citep{pee80a,mar02a}.
If the point process considered is homogeneous and isotropic, the correlation function can be expressed simply in terms of the distance between the points, i.e. $r \equiv | \mathbf{r} |$.
However, this is not the case when studying a sample from a redshift galaxy survey.
Although the galaxy distribution is intrinsically isotropic, the way in which it is measured is not, as the line-of-sight component of each position is derived from the observed redshift.

A way around this issue is the use of the projected correlation function $w_{\rm p}(r_{\rm p})$, first introduced by \citet{dav83a} to deal with the redshift-space effects present in spectroscopic samples \citep{kai87a,ham98a}. 
As shown in \citet{arn09a}, this same approach can be used to deal with samples of photometric redshifts, and we use it in this paper.
In this approach, we first separate the redshift-space distance between any pair of galaxies in two components: parallel ($\pi$) and perpendicular ($r_{\rm p}$) to the line of sight.\footnote{Taking $\mathbf{s}_1$ and $\mathbf{s}_2$ to be the position vectors of the two galaxies, these components are defined as $\pi \equiv \left| \mathbf{s} \cdot \mathbf{l} \right| / \left| \mathbf{l} \right|$ and $r_{\rm p} \equiv \sqrt{\mathbf{s} \cdot \mathbf{s} - \pi^2}$, where $\mathbf{s} \equiv \mathbf{s}_2 - \mathbf{s}_1$, and $\mathbf{l} \equiv  \mathbf{s}_2 + \mathbf{s}_1$.}
We  compute the correlation function as function of these components, $\xi(r_{\rm p},\pi)$, and define the projected correlation function $w_{\rm p}(r_{\rm p})$ as 
\begin{equation}
  \label{eq:1}
  w_{\rm p}(r_{\rm p}) \equiv 2 \int_0^{+\infty} \xi_{\rm s}(r_{\rm p}, \pi) \mathrm{d}\pi \, .
\end{equation}
We estimate $\xi(r_{\rm p},\pi)$ following \citet{lan93a}. 
We first generate an auxiliary random Poisson process following the same selection function as our sample, as defined in Section~\ref{sec:selfunc}. 
We compute, for a given bin in the distance components $(r_{\rm p},\pi)$, the number of pairs in our galaxy catalogue ($DD$), in our random catalogue ($RR$), and the number of crossed pairs between both catalogues ($DR$).
The correlation function is estimated as
\begin{equation}
  \label{eq:2}
  \xi(r_{\rm p},\pi) = 1 + \left(\frac{N_{\rm R}}{N_{\rm D}}\right)^2 \frac{DD(r_{\rm p},\pi)}{RR(r_{\rm p},\pi)} - 2 \frac{N_{\rm R}}{N_{\rm D}} \frac{DR(r_{\rm p},\pi)}{RR(r_{\rm p},\pi)} \, ,
\end{equation}
where $N_{\rm D}$ is the number of galaxies in our sample, and $N_{\rm R}$ is the number of points in the auxiliary random catalogue. 
In this work, we always fix $N_{\rm R} = 20 N_{\rm D}$.
We tested that our results do not change if we increase the number of random points used to  $N_{\rm R} = 50 N_{\rm D}$.

The projected correlation function defined in equation~(\ref{eq:1}) does not depend on the line-of-sight component of the separation $\pi$ and thus, to first order, is not affected by the uncertainty on the photometric redshift determination. 
However, in a real survey, we can not use this definition, as we can not calculate the integral in equation~(\ref{eq:1}) up to infinity. 
We calculate instead
\begin{equation}
  \label{eq:4}
  w_{\rm p}(r_{\rm p}, \pi_{\rm max}) \equiv 2 \int_0^{\pi_{\rm max}} \xi_{\rm s}(r_{\rm p}, \pi) \mathrm{d}\pi \, ,
\end{equation}
which introduces a bias in the result, which is now dependent on the redshift-space effects.
The upper limit $\pi_{\rm max}$ has to be chosen in each case with the aim of minimising this bias, but also of avoiding the introduction of too much additional noise in the calculation.

In Appendix~\ref{sec:pimax} we explore this issue in detail for the case of photometric redshift surveys like ALHAMBRA, using both an analytical model including Gaussian photo-$z$ errors and the full mock catalogues described in Sect.~\ref{ssec:mock-catalogues}.
We study the bias introduced by the finite integration limit, and calculate the minimum value of $\pi_{\rm max}$ needed given the statistical uncertainty in our measurements.
Accounting for this study, we use throughout $\pi_{\rm max} = 200 \hMpc$, which is appropriate for the ALHAMBRA samples considered here.
As a further test, we study the change of our results with $\pi_{\rm max}$ in Appendix~\ref{sec:test-robustness-our}.
Hereafter, we omit the explicit dependence of $w_{\rm p}$ on the value of $\pi_{\rm max}$, and just write $w_{\rm p}(r_{\rm p}) \equiv w_{\rm p}(r_{\rm p}, \pi_{\rm max} = 200 \hMpc)$.

\subsection{Integral constraint}
\label{ssec:integral-constraint}

The integral constraint \citep{pee80a} is a bias in the estimation of the correlation function due to the use of a finite volume.
It is related to the fact that the correlations are measured with respect to the mean density of the sample considered (the particular survey) instead of with respect to the global mean (that of the parent population). 
We can derive the effect of this constraint on $w_{\rm p}$ based on that of the three-dimensional correlation function $\xi$.
When $\xi$ is measured using an estimator such as that of equation~(\ref{eq:2}), it can be shown that the bias introduced by the integral constraint is given, at first order, by \citep{ber02a,lab10a}
\begin{equation}
 \label{eq:3}
  \xi(\mathbf{r}) = \xi^{\rm true}(\mathbf{r}) - K \, ,
\end{equation}
where
\begin{equation}
  \label{eq:5}
  K  \equiv \frac{1}{V^2} \int_V \int_V \mathrm{d}^3 \mathbf{r_1} \mathrm{d}^3 \mathbf{r_2} \xi^{\rm true}(\mathbf{r_2} - \mathbf{r_1}) \, ,
\end{equation}
and $V$ is the volume of the survey.
Using equation~(\ref{eq:4}) this translates into a bias on the estimated projected correlation function $w_{\rm p}(r_{\rm p}, \pi_{\rm max})$ which depends also on $\pi_{\rm max}$,
\begin{equation}
  \label{eq:6}
  w_{\rm p}(r_{\rm p}, \pi_{\rm max}) = w_{\rm p}^{\rm true}(r_{\rm p}, \pi_{\rm max}) - 2K \pi_{\rm max} \, .
\end{equation}

To correct the measured values of $w_{\rm p}$ for the integral constraint using equation~(\ref{eq:6}), one needs to know the true underlying correlation function.
Here we choose an alternative approach, by including the integral constraint correction in the models we fit to the data.
In practice, we follow \citet{Roche1999o} and make use of the auxiliary Poisson catalogue to compute numerically the double integral in equation~(\ref{eq:5}) as 
\begin{equation}
  \label{eq:7}
  K \simeq \frac{\sum_i RR(r_i) \xi^{\rm model}(r_i)}{\sum_i RR(r_i)} = \frac{\sum_i RR(r_i) \xi^{\rm model}(r_i)}{N_R(N_R - 1)} \, ,
\end{equation}
where we use the same notation as in equation~(\ref{eq:2}), and where the sum is over bins in distance extending up to the largest separations in the survey.
In all cases, however, we check that the value of the integral constraint correction is small compared with the errors on $w_{\rm p}$ (as can be seen in Fig.~\ref{fig:wpresults}), so our results are not sensitive to the details of the estimation of $K$.

\subsection{Error estimation}
\label{ssec:error-estimation}

To estimate the statistical error on our $w_{\rm p}(r_{\rm p})$ measurements, we use the standard block bootstrap method \citep[see e.g.][]{nor08a}, making use of the fact that the survey consists of 7 totally independent fields.
We generate $N_{\rm b} = 1000$ bootstrap realisations for each calculation, using the fields as bootstrap regions.
Each of these realisations is created by selecting 7 fields at random, allowing for repetition.
We then compute the projected correlation function for each bootstrap realisation using equations~(\ref{eq:2}) and (\ref{eq:4}).
We obtain the error of $w_{\rm p}$ at each bin in $r_{\rm p}$ as the standard deviation of the measurements from the $N_{\rm b}$ bootstrap realisations.
To account for the covariance between bins in $r_{\rm p}$ when fitting a model to our data, we repeat the $\chi^2$ fitting for the $N_{\rm b}$ realisations, using only the derived diagonal errors.
Our estimate of the error on each model parameter is then the standard deviation of the best values obtained for the $N_{\rm b}$ realisations.

We test in Appendix~\ref{sec:systematics} this error estimation and model fitting procedure for the case of ALHAMBRA using the mock galaxy catalogues described in Sect.~\ref{ssec:mock-catalogues}.
We show that it produces an unbiased estimate of the galaxy bias and of its uncertainty.

We also compared our bootstrap error estimate with the standard jackknife method \citep[see e.g.][]{nor08a}. 
We obtained that the error on $w_{\rm p}(r_{\rm p})$ estimated using both methods is consistent for $r_p \gtrsim 1 \hMpc$.
For $r_{\rm p} \lesssim 1 \hMpc$ the jackknife method slightly underestimates the error with respect to the bootstrap estimate.

%%%%%%%%%%%%%%%%%%%%%%%%%%%%%%%%%%%%%%%%%%%%%%
\section{Correlation functions for ALHAMBRA samples}
\label{sec:results}

\begin{figure*}
  \centering
  \includegraphics[width=\textwidth]{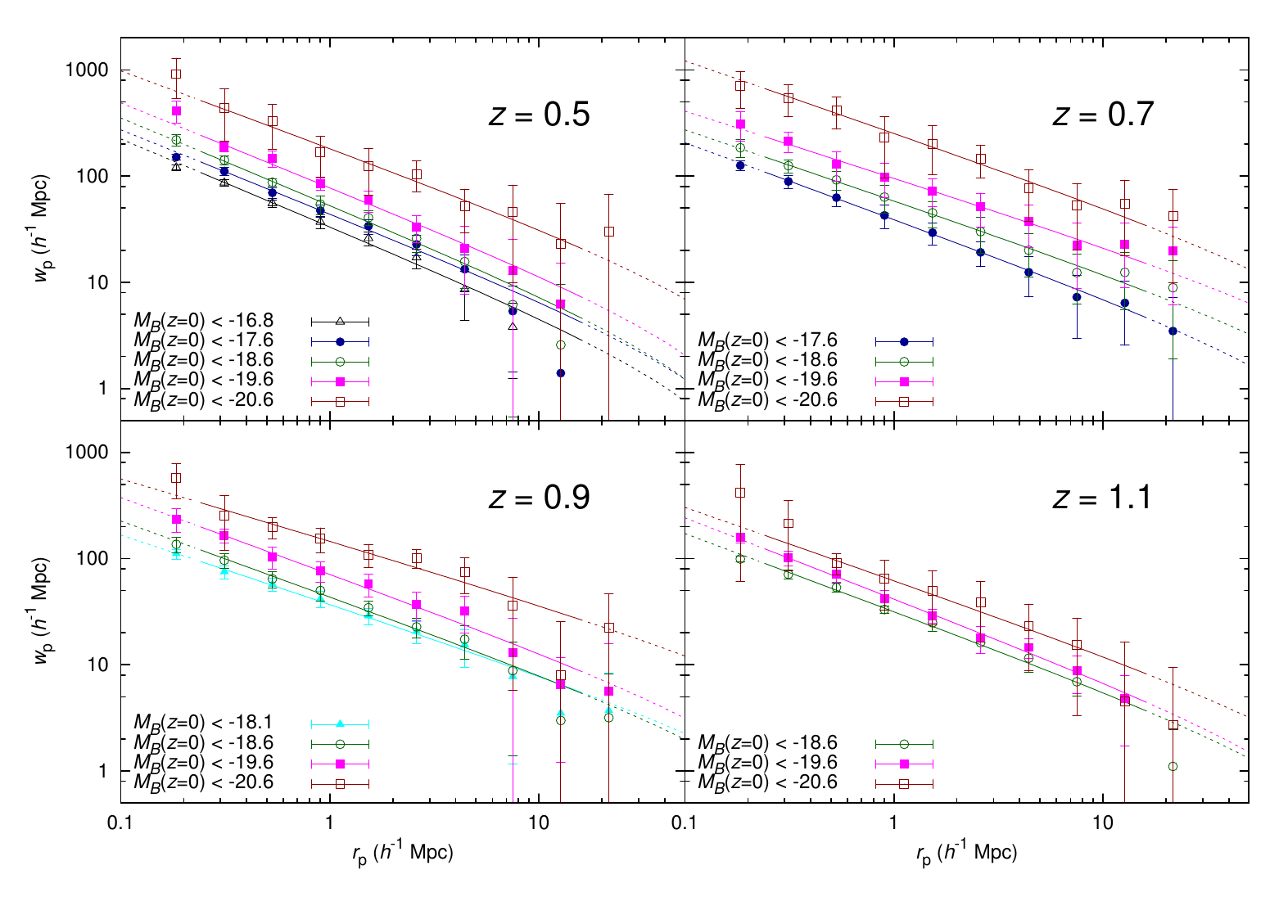}
  \caption{Projected correlation functions for the samples selected in absolute magnitude $M_B$, and redshift (see Table~\ref{tab:samples}). We omit some of the samples for clarity. The solid lines show the corresponding best-fit power laws, according to equation~(\ref{eq:9}), in the range in which the fit was done. Dashed lines show the extrapolation of these models to larger or smaller scales.}
  \label{fig:wpresults}
\end{figure*}

We show the resulting projected correlation functions $w_{\rm p}(r_{\rm p})$ for the different samples selected in redshift and luminosity in Fig.~\ref{fig:wpresults}.
When comparing the results for samples at a given redshift bin we see clearly the effect of segregation by luminosity: bright galaxies are systematically more clustered than faint ones.
This effect can be readily seen in all four redshift bins.
Moreover, we see that all results show approximately a power-law behaviour for scales $r_{\rm p} \gtrsim 0.2 \hMpc$.
We focus here on these scales, and leave the study of smaller scales for a later work.

\subsection{Power-law modelling of the correlation functions}
\label{ssec:powlaw}

In order to study the change of the clustering properties with luminosity and redshift, we fit the obtained projected correlation function $w_{\rm p}(r_{\rm p})$ of each sample using a power law model.
Following the standard practice, we assume the real-space correlation function $\xi(r)$, is given by
\begin{equation}
  \label{eq:8}
  \xi^{\rm pl}(r) = \left( \frac{r}{r_0} \right)^{-\gamma} \, .
\end{equation}
When transforming this model, using equation~(\ref{eq:1}), to a model for $w_{\rm p}(r_{\rm p})$, we also obtain a power law which, expressed in terms of the parameters $r_0$ and $\gamma$ above is given by
\begin{equation}
  \label{eq:wppowlaw}
  w_{\rm p}^{\rm pl}(r_{\rm p}) = r_{\rm p} \left( \frac{r_0}{r_{\rm p}} \right)^{\gamma} \frac{ \Gamma(1/2) \Gamma\left[ (\gamma - 1)/2 \right]}{\Gamma(\gamma/2)} \, ,
\end{equation}
where $\Gamma(\cdot)$ is Euler's Gamma function.
Fitting the power-law model of equation~(\ref{eq:wppowlaw}) to our observed data, we can study the change of both the slope $\gamma$ and the correlation length $r_0$ with the properties of each sample.

In practice, we modify this power-law model by adding the effect of the integral constraint described in Sect.~\ref{ssec:integral-constraint}.
Following equation~(\ref{eq:6}) and leaving explicit the dependence on the model parameters $(r_0, \gamma)$, the model projected correlation function is
\begin{equation}
  \label{eq:9}
  w_{\rm p}^{\rm model}(r_{\rm p}|r_0, \gamma) = w_{\rm p}^{\rm pl}(r_{\rm p} | r_0, \gamma) - 2 K(r_0, \gamma) \pi_{\rm max} \, ,
\end{equation}
where $w_{\rm p}^{\rm pl}(r_{\rm p} | r_0, \gamma)$ is given by equation~(\ref{eq:wppowlaw}), and the integral constraint term $K(r_0, \gamma)$ is obtained from equation~(\ref{eq:7}) using the power-law model for the three-dimensional correlation function of equation~(\ref{eq:8}).
We fit the model of equation~(\ref{eq:9}) to the projected correlation function measured for our different samples in the range $0.2 < r_{\rm p} < 17 \hMpc$.
We obtain the best-fit parameters $\gamma$, $r_0$ in each case using a standard $\chi^2$ minimisation method, and their error using the method described in Sect.~\ref{ssec:error-estimation}.

\begin{figure}
  \centering
  \includegraphics[width=\columnwidth]{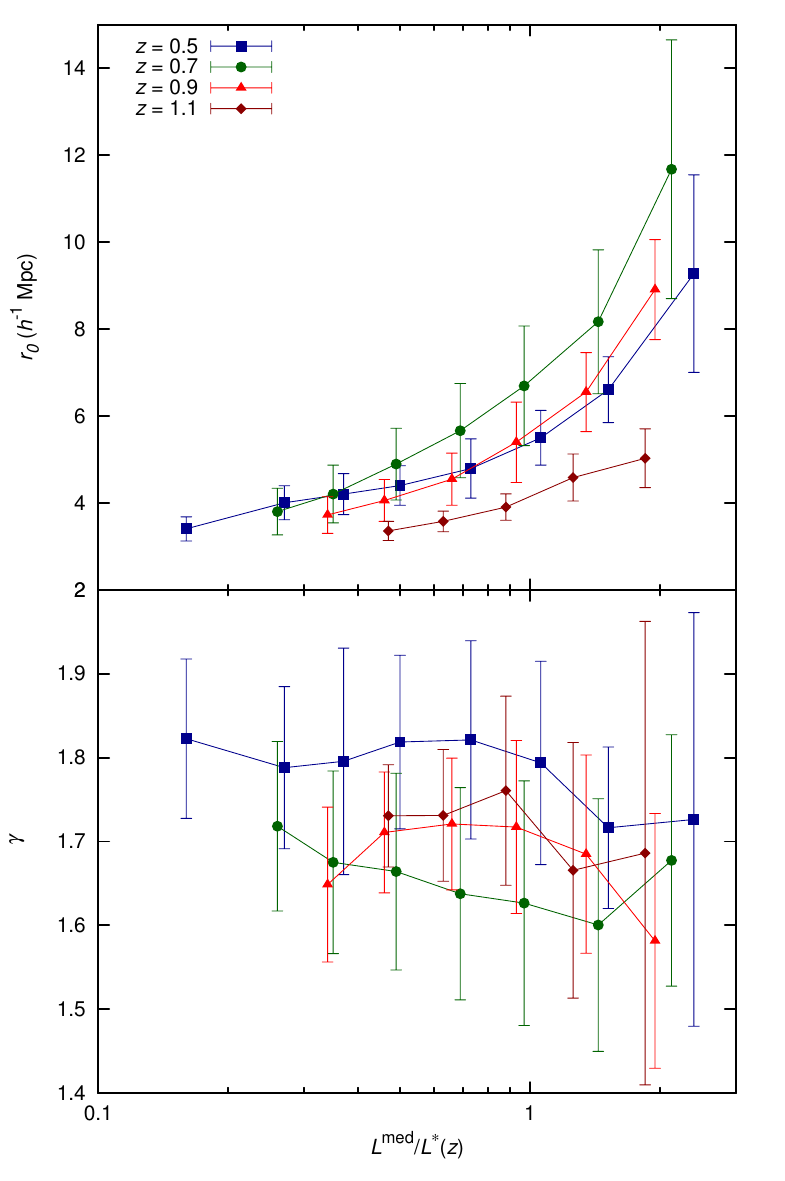}
  \caption{Parameters $r_0$ and $\gamma$ obtained from the power-law fits for the different samples, as a function of the rest-frame $B$-band median luminosity, for each of the redshift bins.}
  \label{fig:powlawparams}
\end{figure}

The best-fit models obtained are shown as solid lines in Fig.~\ref{fig:wpresults}. 
The effect of the integral constrain produces a slight deviation from a straight line (in the log-log plot) at larger scales, very small compared with the errors.
We plot in Fig.~\ref{fig:powlawparams} the resulting parameters $\gamma$, $r_0$ for each of our redshift bins, as function of the median $B$-band luminosity expressed as function of $L^{*}(z)$.

From the bottom panel of Fig.~\ref{fig:powlawparams} we conclude that the slope $\gamma$ is approximately constant, with a value $\gamma \sim 1.75$. This is in agreement with previous studies at similar redshifts \citep{coi06a,mar13a}, although \citet{pol06a} found significantly steeper slopes for the brightest samples.
The results for $r_0$ shown in the top panel of Fig.~\ref{fig:powlawparams}, however, show clear evidence of luminosity segregation, as already observed qualitatively in Fig.~\ref{fig:wpresults}.
In all cases, luminous galaxies are more clustered than faint ones.
However, the change of $r_0$ with redshift is not monotonic.
While the results at $z = 0.5$ and $z = 0.9$ are very similar, the bin at $z = 0.7$ shows a stronger clustering.

The bin at $z = 1.1$ shows a behaviour clearly different to the other three redshift bins. 
On one side, the $r_0$ values for this bin are consistently smaller than those of the lower redshift bins.
On the other side, its dependence on luminosity is much weaker.
However, it is difficult to interpret the results for this last bin, as there is a possible selection bias affecting it.
The reason for this bias is that, for this redshift range, the rest-frame $4000$~\AA\ break is crossing the observer-frame $I$ band used for the selection of our catalogue.
This means that the selection function is changing inside the redshift bin, and in particular this will affect the selection of red passive galaxies (which we expect to show a stronger clustering).
We do not study further this redshift bin in this work, but will study it in more detail in Hurtado-Gil et al. (in prep.), where we focus on the clustering as function of spectral type.

For the three bins at $z \leq 1$, we analyse the clustering properties in detail in the next sections.
First, we separate the evolution of the clustering of the underlying matter density field from that of the bias of our different samples in Sect.~\ref{ssec:bias}.
Then, we study the effect sample variance has on our results, and develop a more robust clustering measurement in Sect.~\ref{sec:cvariance}.

\subsection{Dependence of bias on luminosity and redshift}
\label{ssec:bias}

We study the bias $b$ of our samples by comparing the observed projected correlation function $w_{\rm p}$ for each sample to that of the matter distribution at the corresponding median redshift $w_{\rm p}^{\rm m}$.
We assume a simple linear model, in which bias is constant and independent of scale,
\begin{equation}
  \label{eq:13}
  w_{\rm p}(r_{\rm p}) = b^2 w_{\rm p}^{\rm m}(r_{\rm p}) \, .
\end{equation}
We restrict our study to the bias in the range $1 < r_{\rm p} < 10 \hMpc$, corresponding mainly to the two-halo term of the correlation function.
We leave a more detailed study using the full halo occupation distribution (HOD) formalism \citep{sco01a,ber02b} for a future work.
We note however, that previous works have shown that the value of the bias obtained using our method is consistent to that using the HOD modelling \citep{zeh10a}.

We use a model for $w_{\rm p}^{\rm m}$ based on $\Lambda$CDM and using values of the cosmological parameters consistent with the WMAP7 results \citep{kom11a}.
In particular, we use a normalisation of the power spectrum $\sigma_8 = 0.816$.
We obtain the matter power spectrum at each redshift using the \textsc{Camb} software \citep{lew00a}, including the non-linear corrections of \textsc{Halofit} \citep{smi03a}.
We then Fourier-transform the power spectrum to obtain the real-space correlation function $\xi(r)$ of matter, and finally obtain the projected correlation function using equation~(\ref{eq:1}).
We perform a $\chi^2$ fit to the model in equation~(\ref{eq:13}) as described in Sect.~\ref{ssec:error-estimation}, keeping our model $w_{\rm p}^{\rm m}(r_{\rm p})$ fixed and with $b$ as the only free parameter.
In this case, we use for each sample the value of the integral constraint $K$ obtained from the best-fit power-law model, and correct the observed $w_{\rm p}(r_{\rm p})$ according to equation~(\ref{eq:6}).

\begin{figure}
  \centering
  \includegraphics[width=\columnwidth]{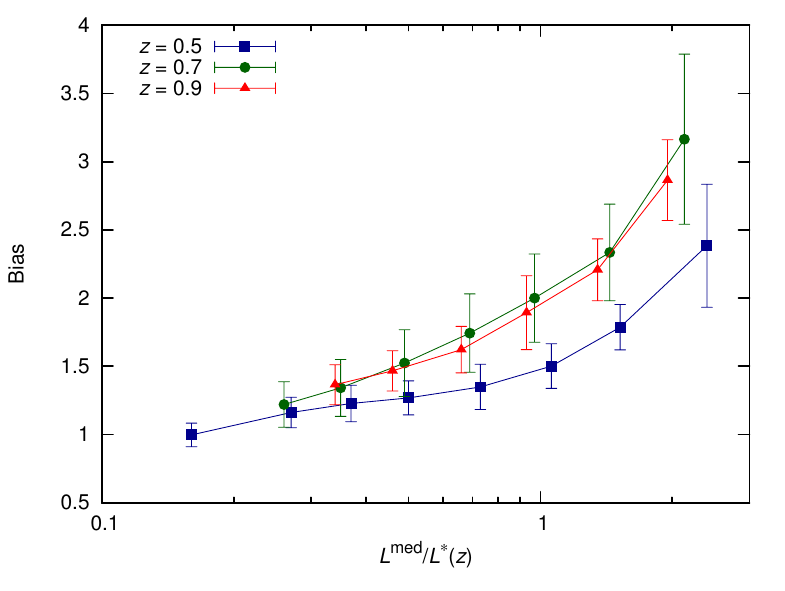}
  \includegraphics[width=\columnwidth]{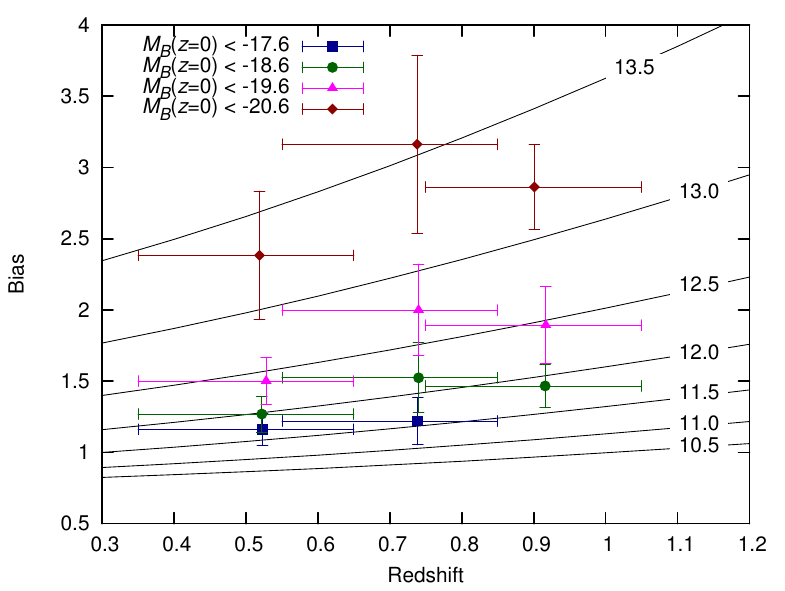}
  \caption{Top: galaxy bias for the different samples from the fit to equation~(\ref{eq:13}), as a function of the median luminosity. Bottom: galaxy bias as function of median redshift for the different luminosity cuts. We omit some of the samples for clarity. The horizontal error bars indicate the full extent of each redshift bin. The solid lines correspond to the bias of haloes above a given mass according to the model of \citet{mo02a}. The label for each of these lines indicates the minimum halo mass in terms of $\log_{10}[M_{\rm h}/(\hMsol)]$.}
  \label{fig:bias}
\end{figure}

The top panel of Fig.~\ref{fig:bias} shows the value of the bias obtained as function of the median luminosity of the sample for each of the three redshift bins considered.
Not surprisingly we see  again the effect of luminosity segregation for all redshift bins, like for  $r_0$ (Fig.~\ref{fig:powlawparams}).
In the bottom panel of Fig.~\ref{fig:bias}, we show the bias as function of redshift for a few of our luminosity-selected samples.
For comparison, we show the bias of haloes of different masses according to the model of \citet{mo02a}.
For the samples with faintest luminosities, the evolution of bias with redshift is not significant.
For the brightest samples, however, the bias does change with redshift.
This evolution is not monotonic, as it seems to have at maximum at $z \sim 0.7$.
Given our uncertainties, this result is not very significant.
However, we study in the next section whether this behaviour is due to the effects of sample variance, and in particular to the contribution of any particular ALHAMBRA field.

\subsection{Analysis of the impact of sample variance on the clustering results}
\label{sec:cvariance}

The use of 7 independent fields in the ALHAMBRA survey is an opportunity to study the effect of sample variance.
Regarding our clustering measurements, we have already used the fact that we have data in several independent fields to estimate the errors in our results, as explained in Sect.~\ref{ssec:error-estimation}.
However, this is based only on a global measure of the variance of the measurements (through the use of the bootstrap technique).

We can go one step further and study the impact of individual fields on our final measurements.
Given the relatively small volume of the survey and, especially, the typical size of the fields, the presence of a large structure in one of the fields could significantly affect our clustering measurement in a given redshift bin.
Similar studies have been performed with other surveys.
For example, when using data from the SDSS, \citet{zeh10a} studied the effect on their results of including or avoiding the SDSS `Great Wall' \citep{Gott2005a}.
\citet{Wolk2013i} performed a similar study for the case of higher-order statistics.

To study the impact of these large structures in our measurements we use the jackknife ensemble fluctuation statistic introduced by \citet{Norberg2011g}.
This statistic is designed as an objective way of identifying `outlier regions': those that, due to the presence of a superstructure, dominate the clustering signal of the whole survey.
In the case of ALHAMBRA, it seems natural to take as jackknife regions our 7 independent fields.
We present here a basic description of this statistic as used in our case for the projected correlation function, but a more detailed description can be found in \citet{Norberg2011g}.

For a given sample, we start by computing the projected correlation function removing from the survey a given field $i$, $w_{\rm p}^i(r_{\rm p})$, and the corresponding rescaled quantity
\begin{equation}
\label{eq:10}
  \Delta_i(r_{\rm p}) = \frac{w_{\rm p}^i(r_{\rm p}) - w_{\rm p}^{\rm full}(r_{\rm p})}{w_{\rm p}^{\rm full}(r_{\rm p})} \, ,
\end{equation}
where $w_{\rm p}^{\rm full}(r_{\rm p})$ refers to the projected correlation function measured from the full sample.
This \emph{jackknife re-sampling fluctuation} $\Delta_i(r_{\rm p})$ therefore quantifies the relative change in $w_{\rm p}$ due to the exclusion of a given field.
To assess the significance of this change, we define the quantity $\sigma_{{\rm tot} - i}^2(r_{\rm p})$ as the rms error of this resampling fluctuations, omitting field $i$,
\begin{equation}
\label{eq:11}
  \sigma_{{\rm tot} - i}^2(r_{\rm p}) = \frac{1}{N_{\rm fields} - 1} \sum_{j \neq i}^{N_{\rm fields} - 1} \Delta_j^2(r_{\rm p}) \, .
\end{equation}
In the case of the ALHAMBRA catalogue used here, $N_{\rm fields} = 7$.
We finally define the \emph{jackknife ensemble fluctuation} $\delta_i$  as the re-sampling fluctuation normalised to its error
\begin{equation}
\label{eq:12}
  \delta_i(r_{\rm p}) = \frac{\Delta_i(r_{\rm p})}{\sigma_{{\rm tot} - i}(r_{\rm p})} \, .
\end{equation}
This is a direct measure of how significant the change in the clustering result for a given sample is when a given field $i$ is either included or excluded.
\citet{Norberg2011g} define an `outlier region' as that for which $\left| \delta_i \right| > 2.5$, where $\delta_i$ is averaged over the range of scales of interest.
We adopt this same limit to define `outlier fields' in the case of ALHAMBRA.
This choice is somehow arbitrary, as full $N$-body simulations would be needed to test the needed value in this case, as done in \citet{Norberg2011g}.

\begin{figure}
  \centering
    \includegraphics[width=\columnwidth]{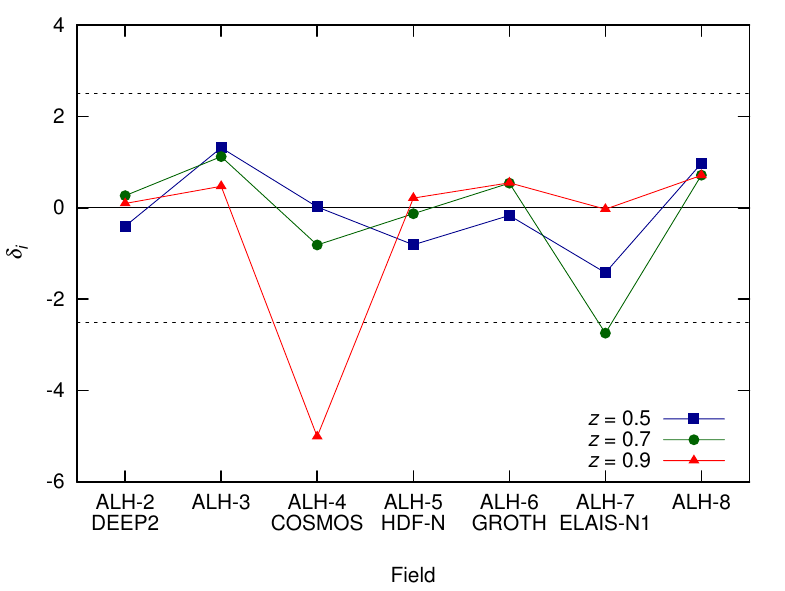}
  \caption{Ensemble fluctuation $\delta_i$ averaged over the range $r_{\rm p} \in [1,10] \hMpc$ for the different redshift bins, as function of the excluded field. These results correspond to the samples selected with $M_B(z=0) < -19.6$, for which $L^{\rm med} \sim L^{*}$. The dashed lines denote our limits $\left|\delta_i\right| = 2.5$ to identify a field as an `outlier'.}
  \label{fig:delta_aver}
\end{figure}

We computed the jackknife ensemble fluctuation $\delta_i$, averaged over the range $1 < r_{\rm p} < 10 \hMpc$ (the same range used to estimate the bias) for the samples selected by $M_B(z=0) < -19.6$ in our three redshift bins, corresponding to $L^{\rm med} \simeq L^*(z)$.
However, as the effects we measure here are due to sample variance, we obtain consistent results when using a different luminosity cut.
We show the results, for the different ALHAMBRA fields, in Fig.~\ref{fig:delta_aver}. 
As expected, in most cases we obtain values $\left|  \delta_i \right| \lesssim 1$ corresponding to the expected variance.
However, we can use the criterion explained above to identify outliers in an iterative way.

The first outlier we identify is the ALH-4 field, for which we obtain the largest value of $\left| \delta_i \right|$, $\delta_i = -5.01$ for the redshift bin centred at $z = 0.9$. 
Once this outlier field is identified, we exclude it from the calculation, and repeat the measurement of $\delta_i$.
Using these new values, we identify an additional outlier: the ALH-7 field, for which we now obtain $\delta_i = -3.45$ for the redshift bin centred at $z = 0.7$.
The original value for this field and redshift bin, when we included also ALH-4 in the calculation, was $\delta_i = -2.74$.
We repeat the process again, excluding both the ALH-4 and ALH-7 fields from the calculation, and find now in all cases values of $\left|\delta_i \right| \leq 1.73$, which we interpret as all fields being equally consistent with each other.

The most obvious outlier is the ALH-4/COSMOS field.
The large negative value of $\delta_i$ obtained means that the inclusion of this field in the survey produces a very significant increase in the measured clustering for this bin.
This is consistent with the fact that previous studies of clustering in the COSMOS survey at similar redshifts have obtained values significantly larger than other similar surveys \citep{McCracken2007a,men09a,tor10a,Skibba2013c}.
The excess clustering  can be explained by the presence of large over-dense structures in this field \citep{Guzzo2007a,Scoville2007d,Kovac2010a}.
In fact, taking into account the particular area covered by the ALH-4 field, we obtain that the four largest structures found by \citet[][see their table~3]{Scoville2007d} are partially included in our sample. 
The central redshifts estimated for these structures are $z = 0.73, 0.88, 0.93, 0.71$, so all of them have substantial overlap with the redshift bin $0.75 < z < 1.05$ where we identify this field as an outlier.
The particularly large over-density of this field is also observed in ALHAMBRA.
The surface density of galaxies is significantly larger in this field than in the rest, as shown in Table~\ref{tab:fields}.
Moreover, the redshift distribution $N(z)$ of this field shows a broad peak centred at $z \sim 0.8$ when compared to the global ALHAMBRA $N(z)$ \citep[see figure 32 in][]{Molino2013a}.

The second `outlier' is the ALH-7/European Large Area ISO Survey North 1 (ELAIS-N1) field. 
Unfortunately, this field is not as well studied as COSMOS and, to the best of our knowledge, there are no previous studies of clustering or identification of large structures at these redshifts. 
However, we also find a peak in the density of clusters and groups  in this field at $z\sim0.7$ using the same ALHAMBRA data set (see Ascaso et al., in prep., for details),  indicating the presence of a large structure at this particular redshift, which could explain the particularly large clustering observed here.

\begin{figure}
  \centering
  \includegraphics[width=\columnwidth]{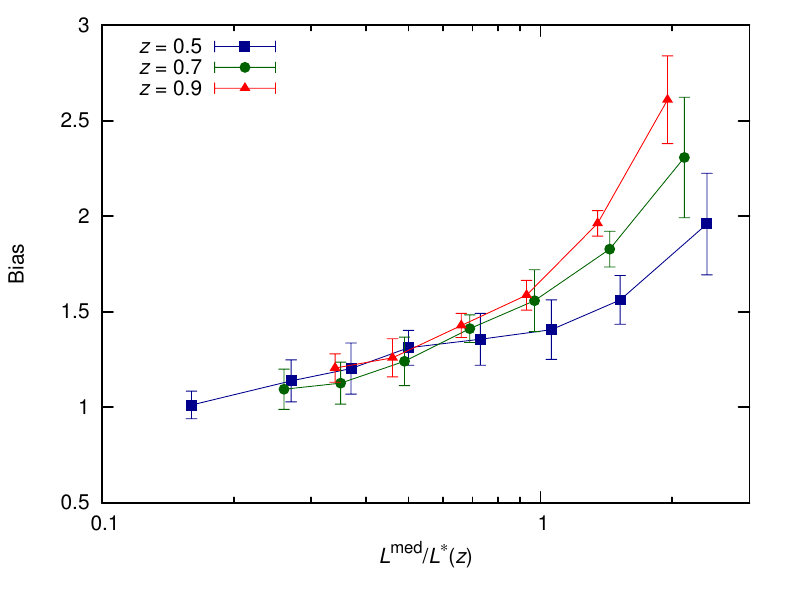}
  \includegraphics[width=\columnwidth]{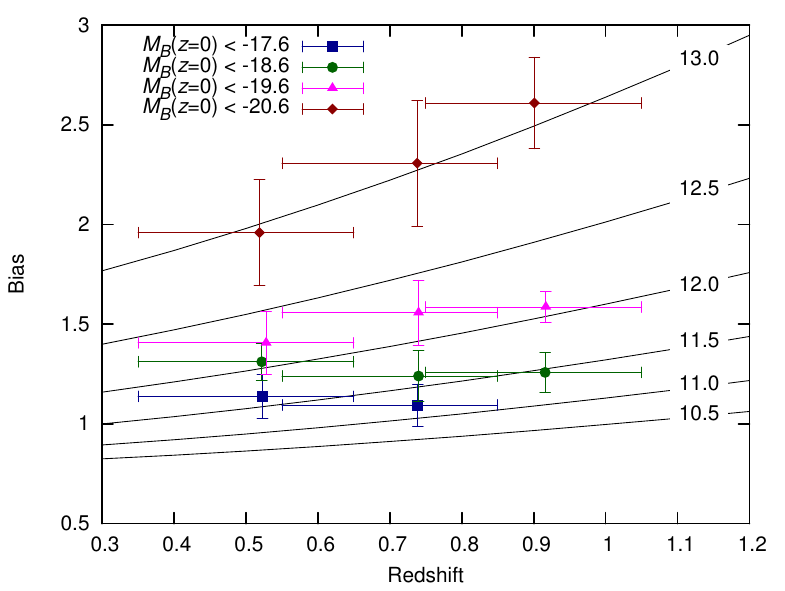}
  \caption{Both panels are identical to these in Fig.~\ref{fig:bias}, for the case in which we totally omit from the calculation the `outlier' fields ALH-4/COSMOS and ALH-7/ELAIS-N1.}
  \label{fig:bias_noalh47}
\end{figure}

Figure~\ref{fig:bias_noalh47} shows the bias of our samples (measured as described in Sect.~\ref{ssec:bias}) as function of their median luminosity and redshift, when we completely omit from the calculation the `outlier fields' ALH-4 and ALH-7. 
We can compare this figure directly to Fig.~\ref{fig:bias}, where we considered the whole survey.
We obtain results very similar to the whole survey for the bin centred at $z = 0.5$. 
This was expected from the results in Fig.~\ref{fig:delta_aver}: the low values of $\left| \delta_i \right|$ for the fields ALH-4 and ALH-7 in this case indicated that removing them would not significantly change the result.
However, we see significant differences for the bins where the removed fields were `outliers', at $z = 0.7$ and $z = 0.9$.
In this case, the bias obtained is smaller now.
The dependence of the bias on luminosity, however, does not change significantly except for the overall normalisation.
This is due to the fact that, for a given redshift bin, we expect sample variance to affect in the same way all the samples regardless of the luminosity selection. 

The error on the bias computed using the bootstrap method has also been greatly reduced.
This was also expected: as we eliminated the greatest outliers, the variance of the remaining measurements is reduced.
However, we note that the original error estimate for the full survey was also affected by the presence of the `outlier' fields, as these imply a very non-Gaussian error distribution.

From the bottom panel of Fig.~\ref{fig:bias_noalh47} we can analyse the evolution of the bias in this case.
For the faintest samples we obtain now an even weaker evolution of the bias.
For the brightest ones we see again a clear variation of bias with redshift, but the observed trend is somewhat different to that seen in Fig.~\ref{fig:bias}.
Now, for our three bins at $z<1$, we see a roughly monotonic trend, with bias increasing with increasing redshift.

Overall, the evolution observed in Fig.~\ref{fig:bias_noalh47} is similar to the bias evolution for haloes above a given mass, according to the model of \cite{mo02a}.
According to that model, the bias we obtain for our different samples correspond to populations of haloes with minimum masses in the range $11.5 \lesssim \log_{10}[M_{\rm h}/(\hMsol)] \lesssim 13.0$.
The bias of galaxies with $L^{\rm med} \simeq L^{*}$ roughly corresponds to that of a halo population with $\log_{10}[M_{\rm h}/(\hMsol)] \gtrsim 12.2$.

\begin{figure}
  \centering
  \includegraphics[width=\columnwidth]{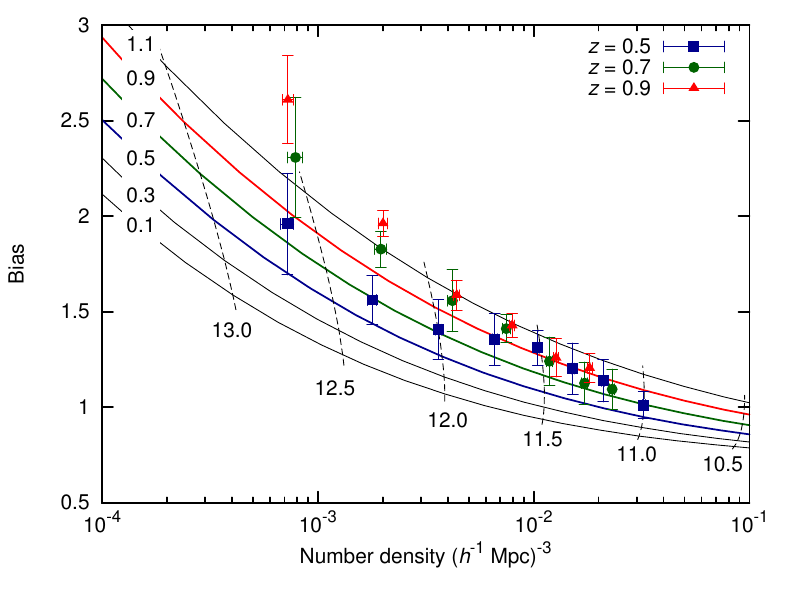}
  \caption{Galaxy bias as a function of the number density of galaxies for our different samples (points). Galaxy bias is obtained from the fit to equation~(\ref{eq:13}), for the case in which we omit the `outlier' fields ALH-4/COSMOS and ALH-7/ELAIS-N1. The lines show the prediction of the model of \citet{mo02a} for haloes above a given mass. Continuous lines show the prediction for fixed values of the redshift (indicated by the labels in the left). Dashed lines correspond to the prediction for fixed values of the minimum halo mass (indicated by the labels in the bottom, in terms of $\log_{10}[M_{\rm h}/(\hMsol)]$). Comparing these predictions for haloes to the observed values, we obtain that the typical mean occupation numbers for the ALHAMBRA galaxies are in the range $\sim 1-3$.}
  \label{fig:ndens_bias}
\end{figure}

To further investigate the relationship between our galaxy samples and the halo populations, we show in Fig.~\ref{fig:ndens_bias} the bias of our samples as a function of their number density.
We compare our results to the prediction for populations of haloes above a given minimum mass from the model of \citet{mo02a}, shown as the continuous (for fixed redshift) and dashed (for fixed minimum mass) lines in the plot.
We can estimate roughly the halo occupation number (i.e. the mean number of galaxies per halo) for a given sample by comparing its number density to that of the halo population at the same redshift and with similar bias.
For the different ALHAMBRA samples, we obtain that the occupation numbers are typically in the range $\sim 1 - 3$, although with a large uncertainty due to our uncertainty in the bias measurement.

\subsection{Comparison to previous results from other surveys}
\label{sec:othersurveys}

We compared our results with previous studies using the largest galaxy surveys to date covering similar redshifts.
\citet{Coupon2012a} studied galaxy clustering in the range $0.2 < z < 1.2$ using data from CFHTLS-Wide\footnote{http://www.cfht.hawaii.edu/Science/CFHLS/}, a broad-band photometric survey covering $\sim 155 \deg^2$.
The bias was derived in each case by fitting a HOD model to the angular correlation function of each sample.
\citet{mar13a} measured the clustering using spectroscopic data from VIPERS\footnote{http://vipers.inaf.it/} covering $\sim 15 \deg^2$, in the range $0.5 < z < 1.1$.
They measured the bias from the measured projected correlation function in the same way as we do here (equation~\ref{eq:13}), and showed that their results were in rough agreement with other (smaller) spectroscopic surveys at similar redshifts such as  DEEP2 \citep{coi06a} and VVDS \citep{pol06a}.
In both cases, the depth of the data used was $i < 22.5$.
We note that the area covered by VIPERS is a subset of that covered by CFHTLS Wide.
The ALH-6 field also overlaps with CFHTLS-Wide.

We also included in this comparison the results in the range $0.5 < z < 1.0$ of \citet{Skibba2013c} using data from PRIMUS.
PRIMUS \citep{Coil2011a} is a survey which uses a low-resolution spectrograph resulting in a typical redshift precision of $\sigma_z/(1+z) = 0.005$ to a depth of $i < 23$.
The data covers five independent fields (including the COSMOS field) covering a total\footnote{This corresponds to the study at $z > 0.5$, where they excluded two additional fields from their analysis. The total area covered by the survey is $9.05 \deg^2$.} of $7.80 \deg^2$.
\citeauthor{Skibba2013c} measured the bias of different samples selected in redshift, luminosity and colour using the projected correlation function in the same way as we describe above (equation~\ref{eq:13}).

\begin{figure}
  \centering
  \includegraphics[scale=1]{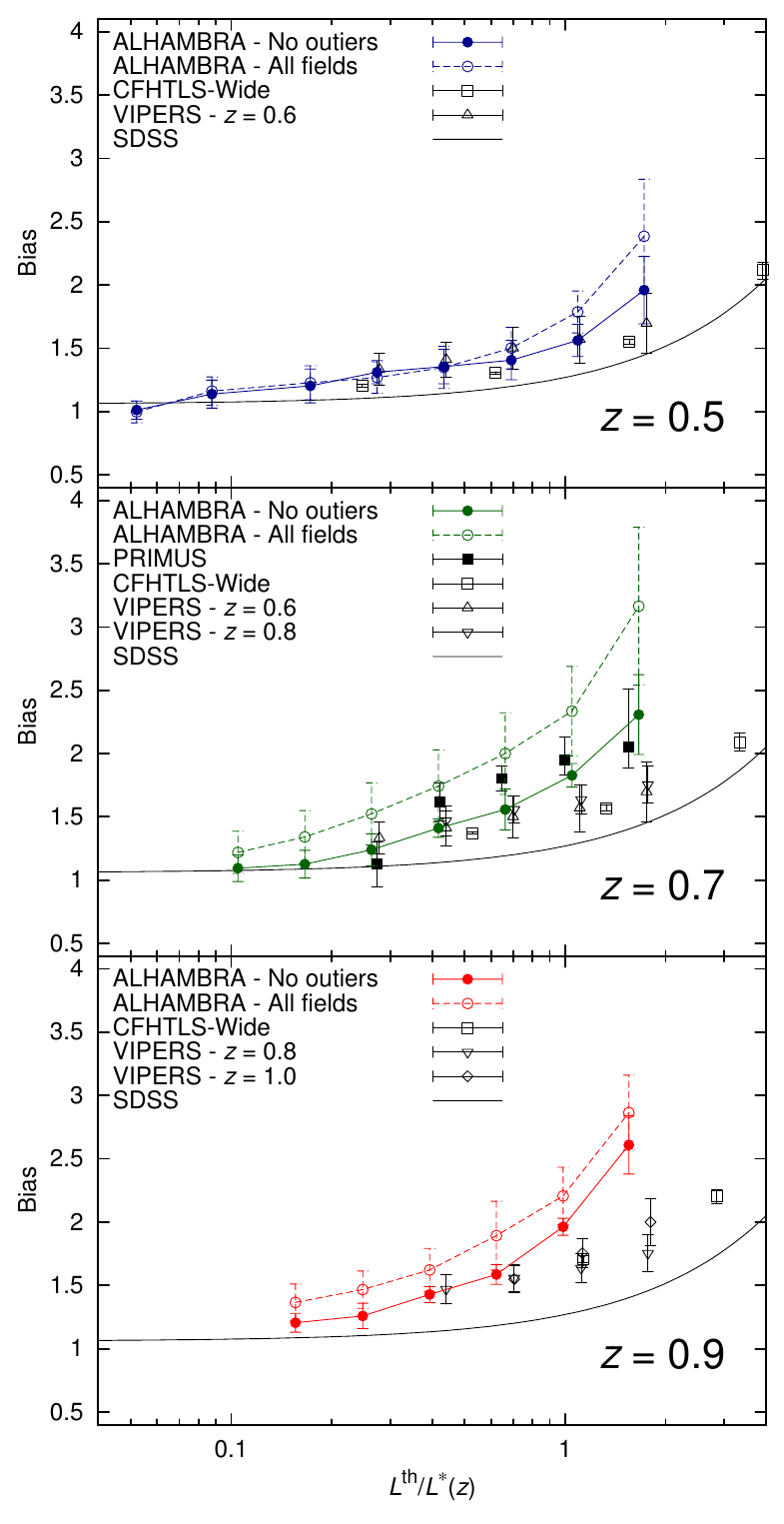}
  \caption{Galaxy bias comparison between ALHAMBRA (this work), VIPERS \citep{mar13a}, CFHTLS-Wide \citep{Coupon2012a} and PRIMUS \citep{Skibba2013c}. 
    The solid line in each panel corresponds to the low redshift SDSS results of \citet{zeh10a}. 
    The bias measurements have been re-normalised to the fiducial value $\sigma_8^{\rm fid} = 0.816$ used in this work.}
  \label{fig:bias_comp}
\end{figure}

In Fig.~\ref{fig:bias_comp} we plot the bias obtained in our different redshift bins as a function of the threshold luminosity $L^{\rm th}$ used to select the different samples in ALHAMBRA, CFHTLS Wide, VIPERS and PRIMUS.
$L^{\rm th}/L^{*}$ is measured at the median redshift of the sample, taking into account the use of different selection parameters $A$ in equation~(\ref{eq:p1alh:Mthevol}).
We note that $L^{\rm th}$ refers to the $B$ band in the case of ALHAMBRA and VIPERS, and to the $g$ band in the case of CFHTLS-Wide and PRIMUS.
In each case, we compare the ALHAMBRA results with the CFHTLS Wide results for the bin centred at the same redshift. 
As \citet{mar13a} used bins centred at different redshifts, we plot in each case the one or two closest bins to the ALHAMBRA one.
In the case of PRIMUS, the actual redshift range of each sample is slightly different with mean redshifts in the range $0.60 - 0.74$, so we plot their results in the central panel.
In each case, we re-normalise the bias by the value of $\sigma_8$ considered.
Changes in bias due to other differences in the cosmology used are much smaller than our errors.
For reference, we also plot as a continuous line the relation derived for low redshifts by \citet{zeh10a} from the SDSS data, which is very similar to that obtained by \citet{nor01a} from the 2dFGRS.
We plot the ALHAMBRA results both for the full survey (dashed lines) and for the case in which we have removed the two `outlier fields' ALH-4 and ALH-7 (solid lines).

We obtain a good agreement between our results and both the CFHTLS-Wide and VIPERS ones, especially considering the significantly smaller area surveyed by ALHAMBRA.
When looking at the $z=0.7$ and $z=0.9$ bins, we see how the result obtained after omitting the outlier fields is in better agreement with the other data than the original results.
This confirms the idea that using the jackknife ensemble fluctuation to identify outlier regions results in a good measurement of the typical clustering properties (bias in this case) of the samples. 
We point out that the comparison presented here was performed only after the full analysis of ALHAMBRA data was finished, so it did not influence the design of the method described in Sect.~\ref{sec:cvariance}.

Our results are also in very good agreement with the PRIMUS results.
We note that PRIMUS obtained slightly larger values of the bias than  CFHTLS-Wide or VIPERS, and they attributed this fact to the presence of the COSMOS field in their sample. 
This is compatible with their results lying between our results with and without the outliers fields included.

We note, however, that the dependence of bias on luminosity appears to be slightly steeper in ALHAMBRA than in previous works.
This is noticeable at the bright end of the $z=0.9$ bin.
It is difficult to assess the significance of this discrepancy, as our bias error estimate is affected by the removal of the `outlier' fields, and the different measurements are highly correlated.
With these caveats in mind, we estimate that the discrepancy for the most extreme case is at the $\lesssim 2 \sigma$ level.
Given its small area, the ALHAMBRA survey is not designed to provide an accurate measurement of low number density samples, nor is the error analysis necessarily adequate for them either. The lowest number density samples (i.e. bright galaxies) require large survey areas to be properly estimated.

Fig.~\ref{fig:bias_comp} shows the complementarity between the different surveys covering this redshift range to study the dependence of galaxy bias on luminosity and redshift.
Large area surveys such as CFHTLS-Wide and VIPERS can measure very accurately the bias of relatively bright samples, $L \gtrsim 0.3 L^{*}(z)$, thus setting the overall normalisation of the $b(L)$ relation at each redshift.
Despite its smaller volume, ALHAMBRA can extend this relation to luminosities $\simeq 1.5 {\rm mag}$ fainter, with our study of the outliers showing that the result is robust to sample variance, except for the overall normalisation.
This larger luminosity range in ALHAMBRA allows us to see clearly the transition from a nearly flat relation at the faint end to a steep one at the bright end.

%%%%%%%%%%%%%%%%%%%%%%%%%%%%%%%%%%%%%%%%%%%%%%
\section{Discussion and conclusions}
\label{sec:conc}

In this work, we have studied the clustering of galaxies in the ALHAMBRA survey and its dependence on luminosity and redshift, in the range $0.35 < z < 1.25$.
To this end, we have used the projected correlation function $w_{\rm p}(r_{\rm p})$, taking into account the uncertainties associated with the use of photometric redshifts, following the method described in \citet{arn09a}. 
We have compared the measured $w_{\rm p}(r_{\rm p})$ to the prediction from our fiducial $\Lambda$CDM model to estimate the bias for the different samples selected in redshift and luminosity.
We also used the method introduced in \citet{Norberg2011g} to study the effect on the clustering measurements of superstructures located in particular ALHAMBRA fields.

The use of the projected correlation function for the case of high-quality photometric redshifts was tested in \citet{arn09a} using a simulated halo catalogue.
Here, we have tested the method using more realistic galaxy mock catalogues (Appendix~\ref{sec:systematics}), and have applied it to real data from the ALHAMBRA survey.
We obtain results that are consistent with larger-area surveys (Sect.~\ref{sec:othersurveys}), and in particular the VIPERS spectroscopic survey, while reaching $1.5$~mag deeper.
This confirms the reliability of the method, and shows that surveys using a large number of medium-band filters can provide very useful data sets for the study of galaxy clustering.
In addition to further results from ALHAMBRA, this indicates good prospects for the planned Javalambre-Physics of the Accelerating Universe Astrophysical Survey\footnote{http://j-pas.org/} \citep[J-PAS,][]{ben08a} and Physics of the Accelerating Universe\footnote{http://www.pausurvey.org/} \citep[PAU,][]{Castander2012} surveys, which will use a similar technique covering larger cosmological volumes.

One of the main characteristics of the ALHAMBRA survey is the mapping of 8 independent fields in the sky (although only 7 are available in the current data set), which provide a useful tool to study the effect of sample variance.
We have studied this issue in two complementary ways. 
On one side, we have used the independence of the fields to obtain a global measure of the clustering uncertainty using the block bootstrap technique described in Sect.~\ref{ssec:error-estimation}.
On the other side, we used the jackknife ensemble fluctuation statistic $\delta_i$ \citep{Norberg2011g} to assess the impact of particular superstructures in the clustering measurements.
This method is based on measuring the clustering omitting one region (field in our case) at a time and comparing it to the global result.
In this way, we have identified the fields ALH-4/COSMOS (at $z \sim 0.9$) and ALH-7/ELAIS-N1 (at $z \sim 0.7$) as `outliers', as the inclusion or omission of each of them changes our results significantly.
We therefore provide also the results for the bias of our samples when we omit these two fields from the calculation, which give a better description of the `typical' clustering properties of the samples, as evidenced by the comparison with the VIPERS and CFHTLS-Wide surveys.

One may want to discuss which is the `correct' result for the bias from this work: that obtained using the full sample (Fig.~\ref{fig:bias}) or that obtained omitting the outlier fields (Fig.~\ref{fig:bias_noalh47}). 
However, it is the combination of both approaches what gives a more complete view of the information about clustering contained in the survey.
On one side, the results obtained after removing the outliers provide information about the typical dependence of galaxy bias on redshift and luminosity.
This is confirmed by the comparison to surveys covering larger volumes, discussed in Section~\ref{sec:othersurveys}.
On the other side, the results for the global sample show how this typical behaviour can be affected by the inclusion or omission of particular fields containing extreme super-structures.
However, the relatively small number of fields covered by ALHAMBRA, and the fact that we only identify either none or one field as an outlier in each of the redshift bins, does not allow us to assess how rare these super-structures are.

Our clustering results give a detailed picture of the dependence of galaxy bias on both luminosity and redshift, summarised in Figs.~\ref{fig:bias_noalh47} and \ref{fig:bias_comp}. 
The depth and photometric redshift reliability of the ALHAMBRA survey allow us to extend the study of the bias to fainter luminosities than previous surveys at similar redshifts.
In this way, the full dependence of bias with luminosity is more clearly seen.
Moreover, our results in Sect.~\ref{sec:cvariance} show that this dependence is reliable, and not significantly affected by sample variance.
At the faint end this relation is nearly flat, up to $L^{\rm med} \simeq L^{*}$ for $z = 0.5$, and up to $L^{\rm med} \simeq 0.5 L^{*}$ for higher redshifts.
At brighter luminosities, the bias increases, following a dependence on $L$ which, for $z = 0.7$ and $z = 0.9$, is significantly steeper than the relation found at low redshift by the SDSS and 2dFGRS surveys.

Regarding the evolution of bias, we see very little dependence of bias with redshift for the faint samples ($L^{\rm med} \lesssim 0.8 L^{*}$), while the evolution is strong for the brighter samples.
In the latter case, for samples with a approximately fixed number density, bias decreases with cosmic time.
This behaviour is consistent with that expected from the halo model, where the bias of the more massive haloes shows much stronger evolution than that of the less massive ones, as illustrated in Figs.~\ref{fig:bias} and \ref{fig:bias_noalh47}.

The comparison of our results with the predicted bias of haloes according to the model of \citet{mo02a} suggests that the galaxies studied reside in haloes covering a range in mass between $\log_{10}[M_{\rm h}/(\hMsol)] \gtrsim 11.5$ (for the samples selected with $M_B(z=0) < -17.6$) and $\log_{10}[M_{\rm h}/(\hMsol)] \gtrsim 13.0$ (for the samples selected with $M_B(z=0) < -20.6$). 
The samples with $L^{\rm med} \simeq L^{*}$ ($M_B(z=0) < -19.6$) are found to correspond to haloes with mass $\log_{10}[M_{\rm h}/(\hMsol)] \gtrsim 12.2$.
From the joint comparison of the bias and number density of our samples to the theoretical prediction for haloes, we obtain that the mean number of galaxies per halo is in the range $\sim 1-3$.

We excluded from this detailed study of the luminosity dependence of the galaxy bias the redshift bin centred at $z = 1.1$. 
As explained in Sect.~\ref{ssec:powlaw}, this is due to the fact that our $I$-band selection could be biasing the sample in that redshift range, affecting in a different way active and passive galaxies.

In this paper, we have focused the study of galaxy clustering in ALHAMBRA on the effect of luminosity segregation and evolution up to $z \sim 1$.
In a companion paper (Hurtado-Gil et al., in prep.) we use this same data set to study the segregation by spectral type in a similar redshift range.
We also plan to extend this work to further redshifts by the use of a NIR-selected catalogue, which will allow us to study the clustering of extremely red objects (EROs, Nieves-Seoane et al., in prep.).

%%%%%%%%%%%%%%%%%%%%%%%%%%%%%%%%%%%%%%%%%%%%%%
\section*{Acknowledgements}

This work is based on observations collected at the German-Spanish Astronomical Center, Calar Alto, jointly operated by the Max-Planck-Institut f\"ur Astronomie (MPIA) and the Instituto de Astrof\'isica de Andaluc\'ia (CSIC).
PAM was supported by an ERC StG Grant (DEGAS-259586).
PN acknowledges the support of the Royal Society through the award of a University Research Fellowship and the European Research Council, through receipt of a Starting Grant (DEGAS-259586).
This work was supported by the Science and Technology Facilities Council (grant number ST/F001166/1), by the Generalitat Valenciana (project of excellence Prometeo 2009/064), by the Junta de Andaluc\'ia (Excellence Project P08-TIC-3531) and by the Spanish Ministry for Science and Innovation (grants  AYA2010-22111-C03-01 and CSD2007-00060). 
This work used the DiRAC Data Centric system at Durham University, operated by the Institute for Computational Cosmology on behalf of the STFC DiRAC HPC Facility (www.dirac.ac.uk). This equipment was funded by BIS National E-infrastructure capital grant ST/K00042X/1, STFC capital grant ST/H008519/1, and STFC DiRAC Operations grant ST/K003267/1 and Durham University. DiRAC is part of the National E-Infrastructure. 

%%BIBLIOGRAPHY PART
\bibliography{ArnalteMur_ALHclustering_v2.bib}

%%APPENDICES%%

\appendix

\section{Optimal value of $\pi_{\rm max}$ for the estimation of the projected correlation function}
\label{sec:pimax}

\subsection{Theoretical determination of the minimum $\pi_{\rm max}$ needed}
\label{sec:analyt-model-determ}

As explained in Sect.~\ref{sec:projcf}, it is not possible to estimate the integral of equation~(\ref{eq:1}) without choosing a finite upper limit $\pi_{\rm max}$, and computing instead $w_{\rm p}(r_{\rm p},\pi_{\rm max})$, as defined in equation~(\ref{eq:4}).
This introduces a bias that has to be accounted for in the modelling.
At the same time, if we extend the measurement to large values of $\pi$ where the signal-to-noise ratio of $\xi(r_{\rm p}, \pi)$ is small, we would be introducing additional noise in the measurement.
In this appendix, we study the relation of this bias with the photo-$z$ errors of the catalogue used, and what is the minimum value of $\pi_{\rm max}$ needed in the case of ALHAMBRA.
To this end, we use the mock catalogues described in Sect.~\ref{ssec:mock-catalogues}, which include photo-$z$ for the galaxies with similar properties to the real data, and a simple analytic model.
In this appendix, we use the cosmological parameters used in the creation of the mocks, $\Omega_{\rm M} = 0.25$, $\Omega_{\Lambda} = 0.75$, $\sigma_8 = 0.9$.

From equations~(\ref{eq:1}) and (\ref{eq:4}), the bias introduced by the finite integration is given by
\begin{equation}
  \label{eq:14}
  \Delta w(r_{\rm p}, \pi_{\rm max}) \equiv w_{\rm p}(r_{\rm p}) - w_{\rm p}(r_{\rm p}, \pi_{\rm max}) = 2 \int_{\pi_{\rm max}}^{+\infty} \xi(r_{\rm p}, \pi) \mathrm{d}\pi \, .
\end{equation}
In principle, given a model for $\xi(r_{\rm p}, \pi)$, one could do the same finite integration in the model, and obtain a prediction directly for $w_{\rm p}(r_{\rm p}, \pi_{\rm max})$. 
However, in this case $w_{\rm p}(r_{\rm p}, \pi_{\rm max})$ is not a real space quantity any longer, and it depends on the way in which redshift space distortions (due to peculiar velocities and photo-$z$) are included in the model.
If we want to avoid this and keep the statistic used as a real space quantity, we should choose a value of $\pi_{\rm max}$ such that the bias $\Delta w(r_{\rm p}, \pi_{\rm max})$ is negligible.
Given the difficulties to model in detail the effect of the photo-$z$ distribution in $\xi(r_{\rm p}, \pi)$, we follow here the latter approach.

We use a galaxy sample selected in redshift and absolute magnitude from the mock catalogues with the limits $0.5 < z_{\rm p} < 0.8$, $M_B -5 \log_{10} h < -17.95$, which is similar to our sample `Z07M1'. 
Following the same method described in Sect.~\ref{sec:projcf}, we obtain the correlation function $\xi(r_{\rm p}, \pi)$ for the 50 realisations, and for the combined $200 \deg^2$ mock. 

We compare the mock results with a simple analytic model obtained using the following steps.
First, we obtain the matter power spectrum $P_{\rm m}(k)$ at the median redshift of the sample using \textsc{Camb} \citep{lew00a}.
We then obtain the real-space galaxy power spectrum $P_{\rm g}(k)$ using a simple HOD model, as described in \citet{abb10a}.
We include the large-scale redshift-space effects following \citet{kai87a} to obtain the redshift-space correlation function $\xi_{\rm s}(r_{\rm p}, \pi)$.
Finally, we include the effect of the photometric redshifts assuming a simple model in which the redshift errors follow a Gaussian distribution.
In this model, the observed correlation function is given by the convolution
\begin{equation}
  \label{eq:15}
  \xi_{\rm phot}(r_{\rm p}, \pi) = \int_{-\infty}^{+\infty} \xi_{\rm s}(r_{\rm p}, \pi')f_{\sigma_{\rm pw}}(\pi - \pi') \mathrm{d}\pi' \, ,
\end{equation}
where $f_{\sigma}(x)$ is the Gaussian distribution of width $\sigma$.
In this case, the width of the distribution is given by the pair-wise photometric redshift uncertainty  $\sigma_{\rm pw} = \sqrt{2}r(\sigma_z)$.
As explained in Sect.~\ref{ssec:samples}, $r(\sigma_z)$ is the comoving separation corresponding to a given photometric redshift uncertainty $\sigma_z$ at the median redshift of the sample.
We choose the HOD parameters of this model (including the bias) to reproduce the observed $w_{\rm p}(r_{\rm p})$ of the mock.

We use both the results from the mock and the analytical model to compute the finite integration bias $\Delta w(r_{\rm p}, \pi_{\rm max})$ defined in equation~(\ref{eq:14}), and find the minimum value of $\pi_{\rm max}$ for which this bias is sufficiently small. 
We express this requirement in terms of the statistical error on $w_{\rm p}(r_{\rm p})$, by requiring the bias to be smaller than 20\% of the estimated statistical uncertainty,
\begin{equation}
  \label{eq:17}
  \frac{\Delta w(r_{\rm p}, \pi_{\rm max})}{w_{\rm p}(r_{\rm p})} \leq 0.2 \sigma_{w_{\rm p}}^{\rm r}(r_{\rm p}) \, ,
\end{equation}
where $\sigma_{w_{\rm p}}^{\rm r}(r_{\rm p})$ is the relative uncertainty in the measurement of $w_{\rm p}(r_{\rm p})$ in a single mock ALHAMBRA realisation obtained from the dispersion of the measurements in the 50 mocks.

\begin{figure}
  \centering
  \includegraphics[width=\columnwidth]{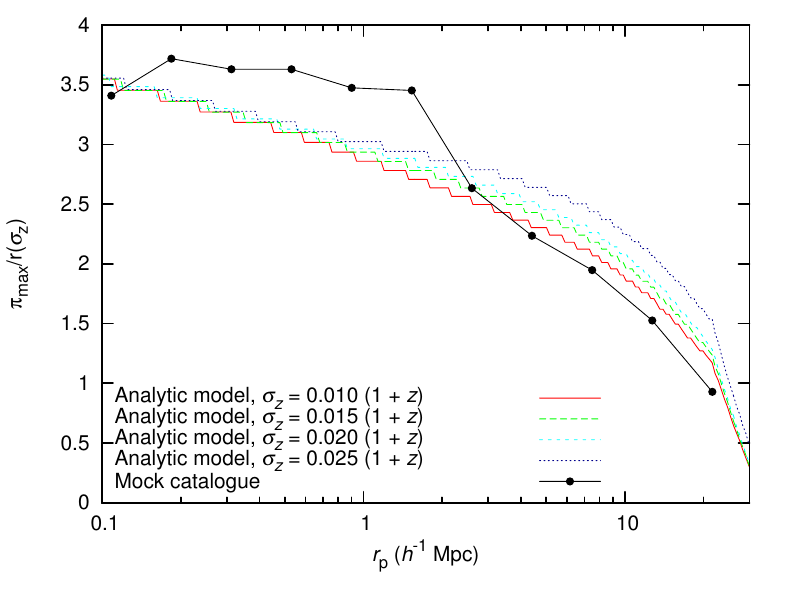}
  \caption{Minimum value of $\pi_{\rm max}$ needed to fulfil our condition (\ref{eq:17}), as function of transverse separation $r_{\rm p}$, for our different models.
  The $\pi_{\rm max}$ in each case is expressed in terms of $r(\sigma_z)$, the comoving separation corresponding to the photometric redshift error $\sigma_z$ of the model, at the median redshift of the bin considered (in this case, $z_{\rm med} \simeq 0.65$). 
  The dashed lines correspond to the theoretical model described in the text, which include the effect of the photometric redshifts using a Gaussian distribution as shown in equation~(\ref{eq:15}). 
  The solid line and points correspond to the measurement in the mock catalogue (combining the 50 ALHAMBRA realisations). 
  For the selected mock sample, we estimate $\sigma_z = 0.025 (1 + z)$, using the method described in Sect.~\ref{ssec:photo-z}.}
  \label{fig:pimax}
\end{figure}

In Fig.~\ref{fig:pimax} we plot the minimum value of $\pi_{\rm max}$ needed to fulfil condition (\ref{eq:17}), as function of $r_{\rm p}$, for our different models: the computation from the combined $200 \deg^2$ mock, and the analytic Gaussian model with different values of the photometric redshift error between $\sigma_z  / (1+z) = 0.010$ and $\sigma_z  / (1+z) = 0.025$.
In the case of the mock catalogue we estimate $\Delta w(r_{\rm p}, \pi_{\rm max})$ using as the reference $w_{\rm p}(r_{\rm p})$ in equation~(\ref{eq:14}) the projected correlation function obtained in real space.
We estimate the typical redshift uncertainty in the mock sample using the BPZ confidence limits in the same way as explained in Sect.~\ref{ssec:photo-z}, in particular including the correction factor of $1.3$, and obtain $\sigma_z / (1+z) = 0.025$.
In all cases, we plot the value of $\pi_{\rm max}$ in terms of $r(\sigma_z)$ for each particular model.

From Fig.~\ref{fig:pimax} we see that overall the required value of $\pi_{\rm max}$ decreases with $r_{\rm p}$. 
This is a consequence of our condition (equation~\ref{eq:17}), given by the fact that the relative error of $w_{\rm p}$ increases with $r_{\rm p}$.
Comparing the different analytical models we see that the different lines are almost coincident for $r_{\rm p} \lesssim 1 \hMpc$, meaning that the required value of $\pi_{\rm max}$ scales linearly with $r(\sigma_z)$.
At larger scales, there is a slight deviation from this proportionality.
Regarding the result obtained from the mock, we see how the non-Gaussianity of the photo-$z$ error distribution has an impact on the observed correlation function $\xi(r_{\rm p}, \pi)$.
This is clearly seen at the smaller scales, $r_{\rm p} \lesssim 2 \hMpc$, where the needed value of $\pi_{\rm max}$ is significantly larger than that predicted by the analytical Gaussian models.
Here, the value of the relative error $\sigma^{\rm r}_{w_{\rm p}}$ is small, so our condition (equation~\ref{eq:17}) is more stringent and the extended wings of the photo-$z$ distribution have the effect of slowing down the convergence of the integral in equation~(\ref{eq:14}).
At larger scales, $r_{\rm p} \gtrsim 2 \hMpc$, our condition (equation~\ref{eq:17}) is much weaker (because $\sigma^{\rm r}_{w_{\rm p}}$ is large), so the details of the wings of the photo-$z$ distribution are less relevant.
Actually, as the mock photo-$z$ distribution is slightly more peaked at the centre than the equivalent Gaussian distribution, we obtain values of $\pi_{\rm max}$ slightly lower than in the analytic case.

Overall, we see that, to fulfil our condition (equation~\ref{eq:17}) over the full range of scales of interest $0.2 < r_{\rm p} < 20 \hMpc$, the minimum value of $\pi_{\rm max}$ needed is $\pi_{\rm max} \simeq 3.5 - 4 r(\sigma_z)$.
This result is in agreement with previous, less detailed estimates \citep{arn09a}.
The particular value of $\pi_{\rm max}$ needed for each ALHAMBRA sample will depend on the details of the correlation function and its error, with the most significant effect being the change in the correlation function error $\sigma^{\rm r}_{w_{\rm p}}(r_{\rm p})$ appearing in equation~(\ref{eq:17}).
As this error depends, at first order, on the sample volume, we repeated the calculation re-scaling it according to the volumes of the actual ALHAMBRA samples used.
We obtained only minor changes in the required value of  $\pi_{\rm max}$ in all cases.
We can therefore estimate the minimum value of $\pi_{\rm max}$ needed for each ALHAMBRA sample from the value of $r(\sigma_z)$ in each case (see Table~\ref{tab:samples}).
Taking $\pi_{\rm max} = 4 r(\sigma_z)$, we obtain values in the range $\pi_{\rm max} \sim 100 - 280 \hMpc$.
As increasing the value of $\pi_{\rm max}$ also introduces additional noise in the measurement, a compromise should be made in deciding the actual value of $\pi_{\rm max}$ to use.
For simplicity, we decided to use a constant value of $\pi_{\rm max}$ for all our samples, fixing it at $\pi_{\rm max} = 200 \hMpc$.
According to the criterion described here, this value of $\pi_{\rm max}$ is adequate for all our samples, except for the faintest samples of each redshift bin.
However, as shown below in Appendix~\ref{sec:test-robustness-our}, we find the bias introduced in these cases to be still acceptable.

\subsection{Test of the robustness of our results with respect to changes in $\pi_{\rm max}$}
\label{sec:test-robustness-our}

\begin{figure}
  \centering
    \includegraphics[width=\columnwidth]{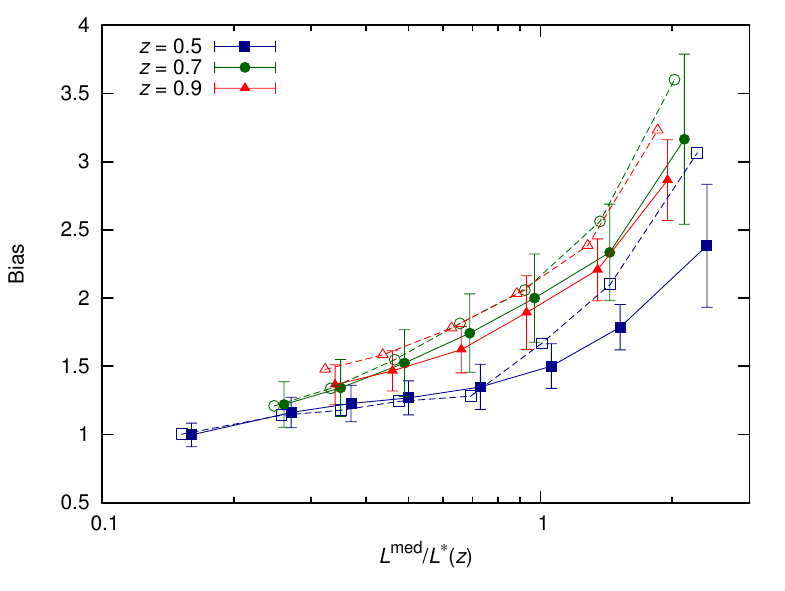}
  \caption{Bias parameter obtained for the different samples from the fit to equation~(\ref{eq:13}), using different values of $\pi_{\rm max}$ in the estimation of the projected correlation function $w_{\rm p}(r_{\rm p})$ (equation~\ref{eq:4}). The solid lines and filled symbols with errorbars correspond to the calculation using $\pi_{\rm max} = 200 \hMpc$ and are the same as shown in the top panel of Fig.~\ref{fig:bias}. The dashed lines and empty symbols (slightly displaced in the horizontal direction for clarity) correspond to the calculation using $\pi_{\rm max} = 350 \hMpc$. We do not show the errorbars in this case, but they are consistently larger than those shown for $\pi_{\rm max} = 200 \hMpc$.}
  \label{fig:bias-pi300}
\end{figure}

We performed an additional test of the effect of our choice of $\pi_{\rm max}$ in our results.
We repeated the calculation of $w_{\rm p}(r_{\rm p})$ for all our samples using a substantially larger value, $\pi_{\rm max} = 350 \hMpc$, in the integration of equation~(\ref{eq:4}).
We did not observe any significant difference in our results but obtained, as expected, larger uncertainties due to the additional noise included in the integration.
As an example of the results obtained in this case, we plot in Fig.~\ref{fig:bias-pi300} the bias as function of luminosity for our samples obtained using different values of $\pi_{\rm max}$.
The solid lines and filled symbols correspond to the results when using $\pi_{\rm max} = 200 \hMpc$, and match the results presented in the top panel of Fig.~\ref{fig:bias}.
The dashed lines and open symbols are our results when we use $\pi_{\rm max} = 350 \hMpc$.
The results in both cases are consistent, especially noting that the errors in the case of using $\pi_{\rm max} = 350 \hMpc$ (not shown in the figure) are consistently larger than those shown for our main results.

\section{Analysis of the reliability of the recovered bias}
\label{sec:systematics}

We used the mock catalogues to test the full process used to estimate the bias of a given galaxy sample and its uncertainty.
We perform this test using the same mock galaxy sample used in Appendix~\ref{sec:analyt-model-determ}, selected in redshift and absolute magnitude as $0.5 < z_{\rm p} < 0.8$, $M_B - 5 \log_{10} h < -17.95$.
We estimated the projected correlation function $w_{\rm p}(r_{\rm p})$ and its error for this sample in each of the 50 mock ALHAMBRA realisations available, following the method described in Sect.~\ref{sec:projcf}.
In the calculation, we used a value of $\pi_{\rm max} = 3 r(\sigma_z)$, as discussed in Appendix~\ref{sec:pimax}.
Given that the redshift uncertainty in this mock sample, $\sigma_z/(1+z) = 0.025$, is somewhat larger than that in the data (see Table~\ref{tab:samples}), this results in a value of $\pi_{\rm max} = 270 \hMpc$, larger than the value used for the data.
However, as we have shown in Appendix~\ref{sec:analyt-model-determ}, the optimal value of $\pi_{\rm max}$ scales with $\sigma_z$, so this value provides an adequate comparison with the calculation done with the data.

\begin{figure}
  \centering
    \includegraphics[width=\columnwidth]{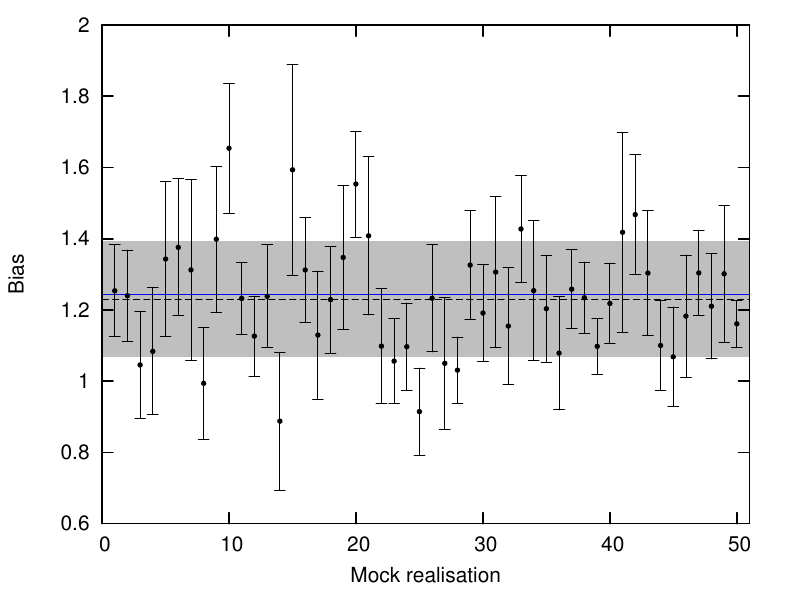}
  \caption{Bias parameter obtained for the galaxy sample defined by $0.5 < z_p < 0.8$, $M_B - 5 \log h < -17.95$ in each of the 50 mock ALHAMBRA realisations. The bias and its uncertainty in each case are estimated using the projected correlation function in the same way as is done for the real data in Sects.~\ref{sec:projcf} and \ref{ssec:bias}. The dashed line shows the mean value obtained from the 50 realisations, $\left\langle b \right\rangle = 1.230$, and the shaded area corresponds to a region of $\pm 1 \sigma_{\rm real}$ around it, where $\sigma_{\rm real} = 0.162$ is the standard deviation of the 50 values. The blue solid line corresponds to the mean bias parameter obtained from the spherically averaged correlation function $\xi(r)$ of the 50 realisations.}
  \label{fig:bias-mockreals}
\end{figure}

We then estimated the bias $b$ and its uncertainty for each realisation using the method described in Sect.~\ref{ssec:bias}, using a model for the matter correlation function matching the Millennium cosmology used in the mocks.
The obtained values are shown in Fig.~\ref{fig:bias-mockreals}. 
The mean value obtained for the bias is $\left\langle b \right\rangle = 1.230$, and the standard deviation of the values from the 50 realisations is $\sigma_{\rm real}(b) = 0.162$.
These values are shown as the dashed line and shaded area in Fig.~\ref{fig:bias-mockreals}.
The values of the uncertainty estimated show a broad distribution, with a mean value of $\left\langle \sigma (b) \right\rangle = 0.159 \pm 0.007$.
This shows that the block bootstrap method used provides an unbiased estimation of the galaxy bias uncertainty.

Finally, we compare these values obtained for the mock catalogues to the real-space result.
To this end, we compute the spherically-averaged real-space correlation function $\xi(r)$ for each of the 50 realisations, using the real position of each galaxy, instead of that estimated from its photometric redshift.
We then obtain the real-space bias using a method analogous to that described in Sect.~\ref{ssec:bias}, using $\xi(r)$ instead of $w_{\rm p}(r_{\rm p})$. 
The mean value of the bias obtained in this way is $\left\langle b_{\rm r} \right\rangle = 1.244 \pm 0.012$ (shown as the continuous blue line in Fig.~\ref{fig:bias-mockreals}).
Therefore, we can estimate the bias in our measurement as $\Delta b =\left\langle b \right\rangle -  \left\langle b_{\rm r} \right\rangle = -0.014$.
This corresponds to $0.09 \sigma_{\rm real}(b)$, and is therefore consistent with the condition imposed in Appendix~\ref{sec:analyt-model-determ} to determine the value of $\pi_{\rm max}$ used (equation~\ref{eq:17}).
We therefore conclude that the method used to recover the real-space clustering (and in particular the galaxy bias) from the ALHAMBRA photometric redshift catalogues using the projected correlation function is reliable, as we recover the direct real-space result within the expected accuracy.

\section{Tables of numerical results}
\label{sec:numer-results}

We present in Table~\ref{tab:numresults} the parameters obtained from the fits of different models to the projected correlation function of our different samples.
The parameters listed are the correlation length $r_0$ and exponent $\gamma$ obtained from the fit to the power-law model in equation~(\ref{eq:9}), and the bias $b$ obtained from the fit to the model in equation~(\ref{eq:13}).
We list both the results obtained using the full survey (with 7 fields) and those obtained when we exclude the `outlier fields' ALH-4/COSMOS and ALH-7/ELAIS-N1 (see Sect.~\ref{sec:cvariance} for details).

\begin{figure}
  \centering
  \includegraphics[width=\columnwidth]{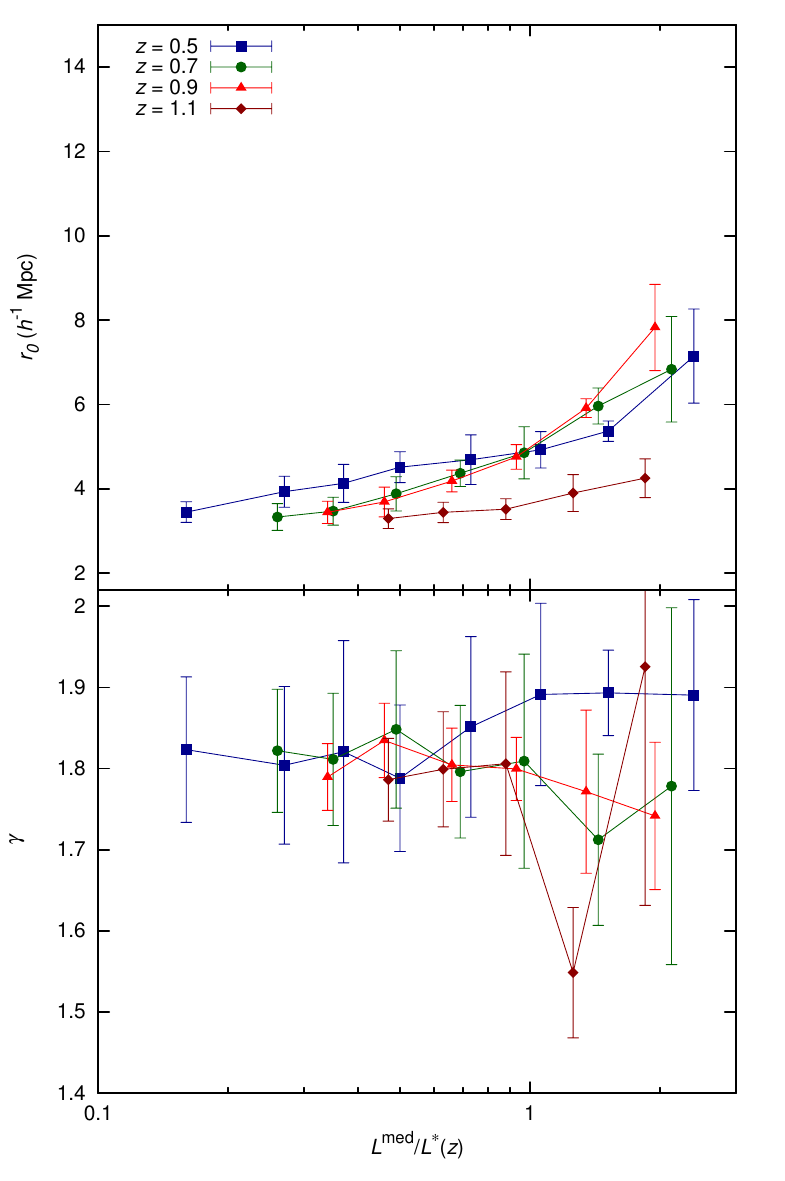}
  \caption{Parameters $r_0$ and $\gamma$ obtained from the power law fits (Sect.~\ref{ssec:powlaw}) for the different samples, as a function of the rest-frame $B$-band median luminosity, for the case in which we omit from the calculation the `outlier' fields ALH-4/COSMOS and ALH-7/ELAIS-N1 (see Sect.~\ref{sec:cvariance}).}
  \label{fig:powlawparams_noout}
\end{figure}

For completeness, we show in Fig.~\ref{fig:powlawparams_noout} the parameters $r_0$ and $\gamma$ obtained from the power-law fits when we exclude the `outlier' fields' from the calculation. This figure can be directly compared to Fig.~\ref{fig:powlawparams}.

\begin{table*}
  \centering
  \caption{Parameters obtained from the fits of different models to the projected correlation function of our different samples. The correlation length $r_0$ and exponent $\gamma$ are obtained from the fit to the power-law model in equation~(\ref{eq:9}) (see Sect.~\ref{ssec:powlaw} for details). The bias is obtained from the fit to the model in equation~(\ref{eq:13}) (see Sect.~\ref{ssec:bias} for details). Results are listed both for the full survey and when excluding the `outlier fields' ALH-4/COSMOS and ALH-7/ELAIS-N1. The sample name listed matches that in Table~\ref{tab:samples}. The results shown in this table correspond to those shown in Figs.~\ref{fig:powlawparams}, \ref{fig:bias}, \ref{fig:bias_noalh47} and \ref{fig:powlawparams_noout}.}
  \label{tab:numresults}
\input{ArnalteMur_table_C1.tex}
\end{table*}

\label{lastpage}

\clearpage

%\label{lastpage}

\end{document}

%% file: ArnalteMur_table_2.tex
%Table recalculated for a 'factor' of 1.3 in the error, instead of the original 1.5
\begin{tabular}{lccrcccccc}
\hline 
Sample	& Redshift range	& $M_B^{\rm th}(0)$ & $N$ & $\bar{n}$ & $z_{\rm med}$	& $M_B^{\rm med}$	& $L^{\rm med}/L^{*}(z_{\rm med})$ & $\sigma_z / (1 + z)$ & $r(\sigma_z)$ \\ 
        &                       &                   &    &  $(10^{-3}\, h^3\,\mathrm{Mpc}^{-3})$ & &                           &         & & $(\hMpc)$ \\ \hline
Z05M0 &  $0.35 - 0.65$  & $-16.8$   &  29496 &  $32.4 \pm 0.5$   & 0.521 &  	-18.32 &    0.16 &      0.0197 &  69.2 \\
Z05M1 &  $0.35 - 0.65$	& $-17.6$   &  19096 &  $21.0 \pm 0.4$   & 0.523 &  	-18.91 &    0.27 &      0.0143 &  50.4 \\
Z05M2 &  $0.35 - 0.65$	& $-18.1$   &  13837 &  $15.2 \pm 0.3$   & 0.524 &  	-19.27 &    0.37 &      0.0133 &  46.5 \\
Z05M3 &  $0.35 - 0.65$	& $-18.6$   &  9530 &   $10.46 \pm 0.22$ & 0.522 &	-19.60 &    0.50 &      0.0135 &  47.4 \\
Z05M4 &  $0.35 - 0.65$	& $-19.1$   &  6012 &   $6.60 \pm 0.17$  & 0.522 &	-19.99 &    0.73 &      0.0131 &  46.1 \\
Z05M5 &  $0.35 - 0.65$	& $-19.6$   &  3295 &   $3.62 \pm 0.12$  & 0.528 &	-20.41 &    1.06 &      0.0086 &  30.2 \\
Z05M6 &  $0.35 - 0.65$	& $-20.1$   &  1627 &   $1.79 \pm 0.08$  & 0.523 & 	-20.80 &    1.52 &      0.0075 &  26.3 \\
Z05M7 &  $0.35 - 0.65$	& $-20.6$   &   657 &   $0.72 \pm 0.05$  & 0.519 &	-21.29 &    2.40 &      0.0068 &  23.8 \\ \hline
Z07M1 &  $0.55 - 0.85$	& $-17.6$   &  33146 &  $23.1 \pm 0.6$   & 0.739 &  	-19.06 &    0.26 &      0.0172 &  61.1 \\
Z07M2 &  $0.55 - 0.85$	& $-18.1$   &  24664 &  $17.2 \pm 0.5$   & 0.740 &  	-19.39 &    0.35 &      0.0139 &  49.3 \\
Z07M3 &  $0.55 - 0.85$	& $-18.6$   &  16979 &  $11.8 \pm 0.4$   & 0.740 &  	-19.75 &    0.49 &      0.0115 &  40.9 \\
Z07M4 &  $0.55 - 0.85$	& $-19.1$   &  10713 &  $7.5 \pm 0.3$    & 0.741 &  	-20.13 &    0.69 &      0.0101 &  35.8 \\
Z07M5 &  $0.55 - 0.85$	& $-19.6$   &  6031 &   $4.20 \pm 0.20$  & 0.740 &	-20.49 &    0.97 &      0.0090 &  32.0 \\
Z07M6 &  $0.55 - 0.85$	& $-20.1$   &  2811 &   $1.96 \pm 0.13$  & 0.740 &	-20.92 &    1.44 &      0.0079 &  27.9 \\
Z07M7 &  $0.55 - 0.85$	& $-20.6$   &  1130 &   $0.79 \pm 0.06$  & 0.738 &	-21.35 &    2.13 &      0.0069 &  24.4 \\ \hline
Z09M2 &  $0.75 - 1.05$	& $-18.1$   &  34712 &  $18.2 \pm 0.5$   & 0.910 &  	-19.51 &    0.34 &      0.0170 &  60.0 \\
Z09M3 &  $0.75 - 1.05$	& $-18.6$   &  24248 &  $12.7 \pm 0.4$   & 0.916 &  	-19.85 &    0.46 &      0.0137 &  48.5 \\
Z09M4 &  $0.75 - 1.05$	& $-19.1$   &  15178 &  $7.94 \pm 0.23$  & 0.916 &  	-20.22 &    0.66 &      0.0115 &  40.6 \\
Z09M5 &  $0.75 - 1.05$	& $-19.6$   &  8413 &   $4.40 \pm 0.15$  & 0.917 &	-20.59 &    0.93 &      0.0103 &  36.3 \\
Z09M6 &  $0.75 - 1.05$	& $-20.1$   &  3830 &   $2.00 \pm 0.09$  & 0.916 &	-21.00 &    1.35 &      0.0089 &  31.5 \\
Z09M7 &  $0.75 - 1.05$	& $-20.6$   &  1387 &   $0.73 \pm 0.04$  & 0.901 &	-21.39 &    1.95 &      0.0083 &  29.5 \\ \hline
Z11M3 &  $0.95 - 1.25$	& $-18.6$   &  23773 &  $10.29 \pm 0.19$ & 1.100 &   	-20.02 &    0.47 &      0.0186 &  65.3 \\
Z11M4 &  $0.95 - 1.25$	& $-19.1$   &  15745 &  $6.82 \pm 0.12$  & 1.108 &  	-20.34 &    0.63 &      0.0152 &  53.1 \\
Z11M5 &  $0.95 - 1.25$	& $-19.6$   &  8677 &   $3.76 \pm 0.07$  & 1.110 &	-20.70 &    0.88 &      0.0130 &  45.5 \\
Z11M6 &  $0.95 - 1.25$	& $-20.1$   &  3868 &   $1.67 \pm 0.05$  & 1.111 &	-21.10 &    1.26 &      0.0114 &  40.0 \\
Z11M7 &  $0.95 - 1.25$	& $-20.6$   &  1285 &   $0.56 \pm 0.02$  & 1.114 &	-21.52 &    1.85 &      0.0103 &  36.0 \\ \hline
\end{tabular}

%% file: ArnalteMur_table_C1.tex
\begin{tabular}{l|ccc|ccc}
\hline 
        & \multicolumn{3}{c}{Fit to the full survey}                   &   \multicolumn{3}{c}{Fit omitting the `outlier' fields ALH-4 and ALH-7} \\
Sample  & $r_0 (\hMpc)$       & $\gamma$           & Bias              & $r_0 (\hMpc)$    & $\gamma$           & Bias \\ \hline
Z05M0   & $3.4  \pm  0.3   $   & $1.82  \pm  0.09   $  & $1.00  \pm  0.09 $ & $3.45\pm 0.25$ & $1.82  \pm  0.09 $ & $1.01 \pm 0.07    $      \\
Z05M1   & $4.0  \pm  0.4   $   & $1.79  \pm  0.10   $  & $1.16  \pm  0.11 $ & $3.9 \pm 0.4 $ & $1.80  \pm  0.10 $ & $1.14 \pm 0.11	     $     \\
Z05M2   & $4.2  \pm  0.5   $   & $1.80  \pm  0.14   $  & $1.23  \pm  0.13 $ & $4.1 \pm 0.5 $ & $1.82  \pm  0.14 $ & $1.20 \pm 0.13     $     \\  
Z05M3   & $4.4  \pm  0.5   $   & $1.82  \pm  0.10   $  & $1.27  \pm  0.12 $ & $4.5 \pm 0.4 $ & $1.79  \pm  0.09 $ & $1.31 \pm 0.09    $     \\  
Z05M4   & $4.8  \pm  0.7   $   & $1.82  \pm  0.12   $  & $1.35  \pm  0.17 $ & $4.7 \pm 0.6 $ & $1.85  \pm  0.11 $ & $1.36 \pm 0.14       $     \\  
Z05M5   & $5.5  \pm  0.6   $   & $1.79  \pm  0.12   $  & $1.50  \pm  0.16 $ & $4.9 \pm 0.4 $ & $1.89  \pm  0.11 $ & $1.41 \pm 0.16     $     \\  
Z05M6   & $6.6  \pm  0.8   $   & $1.72  \pm  0.10   $  & $1.79  \pm  0.17 $ & $5.37\pm 0.24$ & $1.89  \pm  0.05 $ & $1.56 \pm 0.13     $     \\  
Z05M7   & $9.3  \pm  2.3   $   & $1.73  \pm  0.25   $  & $2.4   \pm  0.5  $ & $7.1 \pm 1.1 $ & $1.89  \pm  0.12 $ & $2.0  \pm 0.3     $     \\   \hline
Z07M1   & $3.8  \pm  0.5   $   & $1.72  \pm  0.10   $  & $1.22  \pm  0.17 $ & $3.3 \pm 0.3 $ & $1.82  \pm  0.08 $ & $1.09 \pm 0.11     $     \\  
Z07M2   & $4.2  \pm  0.7   $   & $1.68  \pm  0.11   $  & $1.34  \pm  0.21 $ & $3.5 \pm 0.3 $ & $1.81  \pm  0.08 $ & $1.13 \pm 0.11     $     \\  
Z07M3   & $4.9  \pm  0.8   $   & $1.66  \pm  0.12   $  & $1.52  \pm  0.24 $ & $3.9 \pm 0.4 $ & $1.85  \pm  0.10 $ & $1.24 \pm 0.13	     $     \\  
Z07M4   & $5.7  \pm  1.1   $   & $1.64  \pm  0.13   $  & $1.74  \pm  0.29 $ & $4.4 \pm 0.3 $ & $1.80  \pm  0.08 $ & $1.41 \pm 0.07    $     \\  
Z07M5   & $6.7  \pm  1.4   $   & $1.63  \pm  0.15   $  & $2.0   \pm  0.3  $ & $4.9 \pm 0.6 $ & $1.81  \pm  0.13 $ & $1.56 \pm 0.16     $     \\  
Z07M6   & $8.2  \pm  1.7   $   & $1.60  \pm  0.15   $  & $2.3   \pm  0.4  $ & $6.0 \pm 0.4 $ & $1.71  \pm  0.11 $ & $1.83 \pm 0.09    $     \\  
Z07M7   & $11.7 \pm  3.0 $     & $1.68  \pm  0.15 $    & $3.2   \pm  0.6  $ & $6.8 \pm 1.3 $ & $1.78  \pm  0.22 $ & $2.3  \pm 0.3     $     \\  \hline
Z09M2   & $3.7  \pm  0.4   $   & $1.65  \pm  0.09   $  & $1.37  \pm  0.15 $ & $3.4 \pm 0.3 $ & $1.79  \pm  0.04 $ & $1.20 \pm 0.07    $     \\  
Z09M3   & $4.1  \pm  0.5   $   & $1.71  \pm  0.07   $  & $1.47  \pm  0.15 $ & $3.7 \pm 0.4 $ & $1.83  \pm  0.05 $ & $1.26 \pm 0.10    $     \\  
Z09M4   & $4.6  \pm  0.6   $   & $1.72  \pm  0.08   $  & $1.62  \pm  0.17 $ & $4.2 \pm 0.3 $ & $1.80  \pm  0.05 $ & $1.43 \pm 0.06    $     \\  
Z09M5   & $5.4  \pm  0.9   $   & $1.72  \pm  0.10   $  & $1.9   \pm  0.3  $ & $4.8 \pm 0.3 $ & $1.80  \pm  0.04 $ & $1.59 \pm 0.08    $     \\  
Z09M6   & $6.6  \pm  0.9   $   & $1.69  \pm  0.12   $  & $2.21  \pm  0.23 $ & $5.91\pm 0.22$ & $1.77  \pm  0.10 $ & $1.96 \pm 0.07    $     \\  
Z09M7   & $8.9  \pm  1.2   $   & $1.58  \pm  0.15   $  & $2.9   \pm  0.3  $ & $7.8 \pm 1.0 $ & $1.74  \pm  0.09 $ & $2.61 \pm 0.23     $     \\  \hline
Z11M3   & $3.36  \pm  0.22 $   & $1.73  \pm  0.06   $  & ---                & $3.29\pm 0.23$ & $1.79  \pm  0.05 $ & ---     \\  
Z11M4   & $3.58  \pm  0.24 $   & $1.73  \pm  0.08   $  & ---                & $3.44\pm 0.24$ & $1.80  \pm  0.07 $ & ---     \\  
Z11M5   & $3.9  \pm  0.3   $   & $1.76  \pm  0.11   $  & ---                & $3.52\pm 0.24$ & $1.81  \pm  0.11 $ & ---     \\  
Z11M6   & $4.6  \pm  0.5   $   & $1.67  \pm  0.15   $  & ---                & $3.9 \pm 0.4 $ & $1.55  \pm  0.08 $ & ---     \\  
Z11M7   & $5.0  \pm  0.7   $   & $1.7  \pm  0.3   $    & ---                & $4.3 \pm 0.5 $ & $1.9   \pm  0.3  $ & ---     \\  \hline
\end{tabular}           
                                             